\documentclass[lettersize,journal]{IEEEtran}
\usepackage{cite}
\usepackage{amsmath,amssymb,amsfonts}
\usepackage{graphicx}
\usepackage{textcomp}
\usepackage{xcolor}
\usepackage{float}
\usepackage[ruled,vlined]{algorithm2e}
\usepackage{algpseudocode}
\usepackage{makecell}
\usepackage{caption}
\usepackage{subfigure}
\usepackage{mdwlist}
\usepackage{enumitem} 

\usepackage{multirow}

\usepackage{titlesec}
\titlespacing*{\paragraph}{0pt}{1ex plus 1ex minus .2ex}{0pt}

\begin{document}

%\title{Hardware-assisted Ransomware Detection and Mitigation using Machine Learning}

\title{Hardware-Assisted Multi-Stage Ransomware Detection with Adaptive Machine Learning}
%\title{Robust and Efficient Ransomware Detection using Hardware-assisted Machine Learning}

%\author{Salma Aboelmagd, Md. Rakibul Ahasan, Ziyu Shu, Zhixin Pan, and Abdulrahman Takiddin}

\author{
\IEEEauthorblockN{Salma Aboelmagd\IEEEauthorrefmark{1}, 
Md. Rakibul Ahasan\IEEEauthorrefmark{1}, 
Zhixin Pan\IEEEauthorrefmark{1}, 
Abdulrahman Takiddin\IEEEauthorrefmark{1}, 
and Ziyu Shu\IEEEauthorrefmark{2}}\\
\IEEEauthorblockA{\IEEEauthorrefmark{1}
Florida State University, Tallahassee, FL, USA\\
\IEEEauthorrefmark{2}
Stony Brook University, Stony Brook, NY, USA}
\thanks{Salma Aboelmagd and Md. Rakibul Ahasan are Graduate Student Members, IEEE.  
Zhixin Pan, Abdulrahman Takiddin, and Ziyu Shu are Members, IEEE.}
}

\markboth{Journal of \LaTeX\ Class Files,~Vol.~14, No.~8, August~2021}%
{Shell \MakeLowercase{\textit{et al.}}: A Sample Article Using IEEEtran.cls for IEEE Journals}


\maketitle

\begin{abstract}

Ransomware is a fast-escalating privacy threat causing significant financial and data losses worldwide. As attacks grow more sophisticated, traditional static signatures and dynamic behavior analysis struggle to defend. In this paper, we  propose an novel framework that couples software-level scanning with hardware-level microprocessor activity monitoring to achieve efficient ransomware detection and mitigation. The contributions are threefold: (1) we employ adversarial training to harden static analysis against obfuscation and leverage hardware-assisted monitoring for real-time detection; (2) we integrate neural architecture search (NAS) to automatically tailor models for stronger generalizability; and (3) We validate the proposed approach through extensive experiments, demonstrating substantially improved detection accuracy and reduced latency compared with existing methods. Moreover, it provides trace-level interpretability that explains the underlying decision process.

\end{abstract}

% \begin{IEEEkeywords}
% Hardware Security, Malware Detction, Machine Learning, Transformer
% \end{IEEEkeywords}

% \fi
\IEEEpeerreviewmaketitle


\section{Introduction}
\label{sec:intro}

Cybersecurity has remained a major concern since the emergence of networked computing systems. Among various security threats, \textit{ransomware}, a special type of malicious software that encrypts critical data and demands payment for its release, has become one of the most destructive and financially damaging forms of cyberattacks. Its growing sophistication has resulted in significant economic losses, operational malfunction, and exposure of sensitive information worldwide. As shown in a recent study~\cite{mcintosh2021ransomware}, ransomware attacks are increasingly targeting large organizations, with estimated global damages approaching \$6~trillion annually. Moreover, attackers have adopted advanced obfuscation and evasion strategies, rendering conventional defenses such as signature-based detection and fuzzing ineffective, highlighting the need for more adaptive detection and mitigation approaches.

\begin{figure}[htbp]
\centering
\vspace{-0.1in}
\includegraphics[width = 0.5\textwidth]{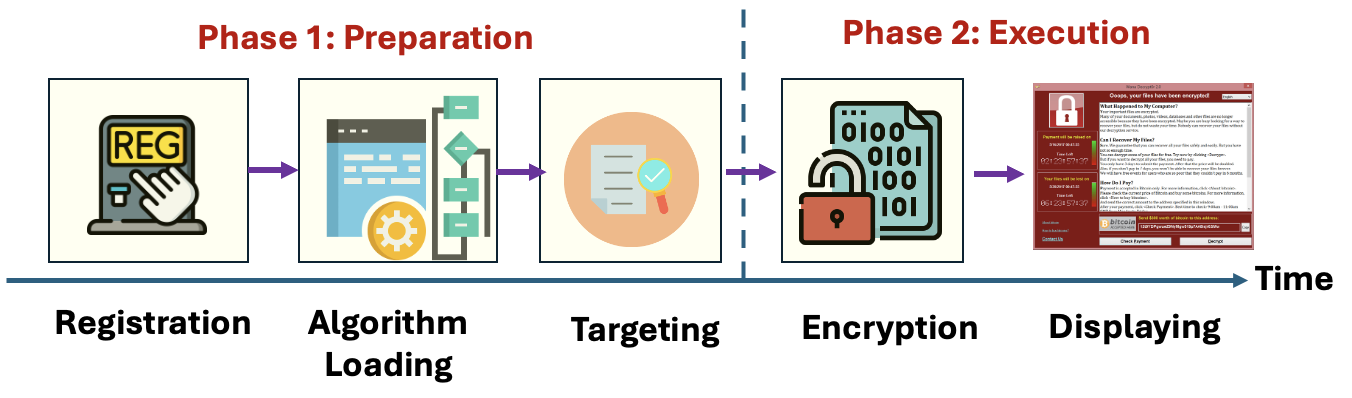}
\caption{Typical ransomware infection workflow consisting of two major phases.}
\vspace{-0.1in}
\label{fig:emp}
\end{figure}

Figure~\ref{fig:emp} illustrates a typical ransomware infection timeline, which generally consists of two major stages. The \textit{preparation phase} involves initialization, loading encryption algorithms, and scanning the system for valuable files. The \textit{execution phase} then rapidly performs destructive actions depending on the ransomware type: (i) Crypto (file encryption with ransom demand), (ii) Locker (blocking system access), (iii) Scareware (displaying fake alerts), or (iv) Hybrid (combining multiple attacks).

Existing ransomware detection methods can be broadly categorized into \textit{static analysis} and \textit{dynamic analysis} approaches. Static analysis inspects executable files or code structures before execution, typically through signature matching or feature recognition. While efficient, such techniques often fail against obfuscated or polymorphic malware, resulting in high false positive rates~\cite{mcintosh2021ransomware, olaimat2021ransomware}. Dynamic or behavioral analysis, instead, monitors runtime activities such as file operations and system calls to identify malicious patterns~\cite{arabo2020detecting}. Although generally more accurate, these approaches are often platform-dependent and suffer from latency, while ransomware protection is a time-critical task. Furthermore, most behavior-based systems produce binary detection results without explaining the underlying reasoning, making it hard to perform post-detection actions like file restoration or infection localization.

% To overcome these challenges, this work proposes a \textit{hardware-assisted, machine learning-based ransomware detection framework} that integrates both static and dynamic analysis in a unified architecture. Unlike prior hybrid methods that rely on manual rule design or expert intervention to reconcile conflicting results, our approach introduces several distinctive contributions:

% \begin{itemize}
%     \item \textbf{Multi-stage detection:} A hierarchical detection pipeline integrates static feature inspection with dynamic monitoring, activating runtime observation only when static indicators suggest potential threats, thereby minimizing overhead and latency.
%     \item \textbf{Automated model optimization:} A neural architecture search (NAS) algorithm automatically configures the machine learning models to maximize detection accuracy and generalizability across ransomware variants.
%     \item \textbf{Model explainability:} The framework incorporates explainable machine learning techniques to attribute features and visualize decision pathways, enhancing transparency and analyst trust.
% \end{itemize}
To overcome these challenges, this work proposes a \textit{hardware-assisted, machine learning (ML) based ransomware detection framework} that integrates static analysis with dynamic runtime monitoring in a unified architecture. 
Unlike prior hybrid approaches that rely on fixed rules or manual intervention to balance accuracy and overhead, our framework is centered on adaptive hardware-assisted monitoring and makes the following contributions:
\begin{itemize}
    
    \item \textbf{Multi-stage detection pipeline:} 
     We propose a hierarchical detection pipeline that integrates static inspection with dynamic hardware-assisted analysis, activating fine-grained runtime monitoring only when necessary to reduce detection latency and system overhead.
    
    \item \textbf{Dynamic hardware-assisted monitoring:} 
    We develop an adaptive runtime monitoring strategy that combines coarse-grained observation with selective fine-grained runtime tracing, enabling low-overhead detection with timely confirmation under obfuscated ransomware behaviors.

    \item \textbf{Adaptive ML optimization:} 
    A neural architecture search (NAS) mechanism automatically configures learning models for both static and dynamic stages, improving robustness and generalizability across diverse ransomware variants.
\end{itemize}

The remainder of this paper is structured as follows. Section~\ref{sec:bgd} reviews the related studies. Section~\ref{sec:prop} presents the proposed detection framework. Section~\ref{sec:exp} discusses experimental evaluations, and Section~\ref{sec:con} concludes the paper.

\section{Background and Related Work}
\label{sec:bgd}
% ML-based Ransomware Detection
% 3 categories
% A picture of the delemma
% A histogram showing generalizability is bad

The research on ransomware detection has gained growing attention in recent years. As discussed in Section~\ref{sec:intro}, the majority of existing methods can be broadly categorized into two types: \textit{static analysis} and \textit{dynamic analysis}.

\textbf{Static analysis} is conducted before program execution, where the executable or source code is examined to identify malicious intent. For instance,~\cite{al2019crypto,cuzzocrea2018novel,faris2020optimizing,feng2017poster,deng2024ransomware} utilized deep file–scanning and code–inspection techniques to detect potential ransomware, while~\cite{alzahrani2018randroid,cimitile2018talos,karimi2017android} explored linguistic analysis of program code for early-stage detection. However, static approaches lack the behavioral information from runtime, which limits their ability. Moreover, certain benign applications, such as disk–encryption tools, may share similar binary patterns with ransomware, causing high false positive rates (FPRs). Additionally, many static methods rely heavily on signature-based pattern matching; when attackers could intentionally disguise or obfuscate code~\cite{lee2023hiding}, these methods may fail to detect the threat accurately.

\begin{figure}[htbp]
\centering
\vspace{-0.1in}
\includegraphics[width = 0.4\textwidth]{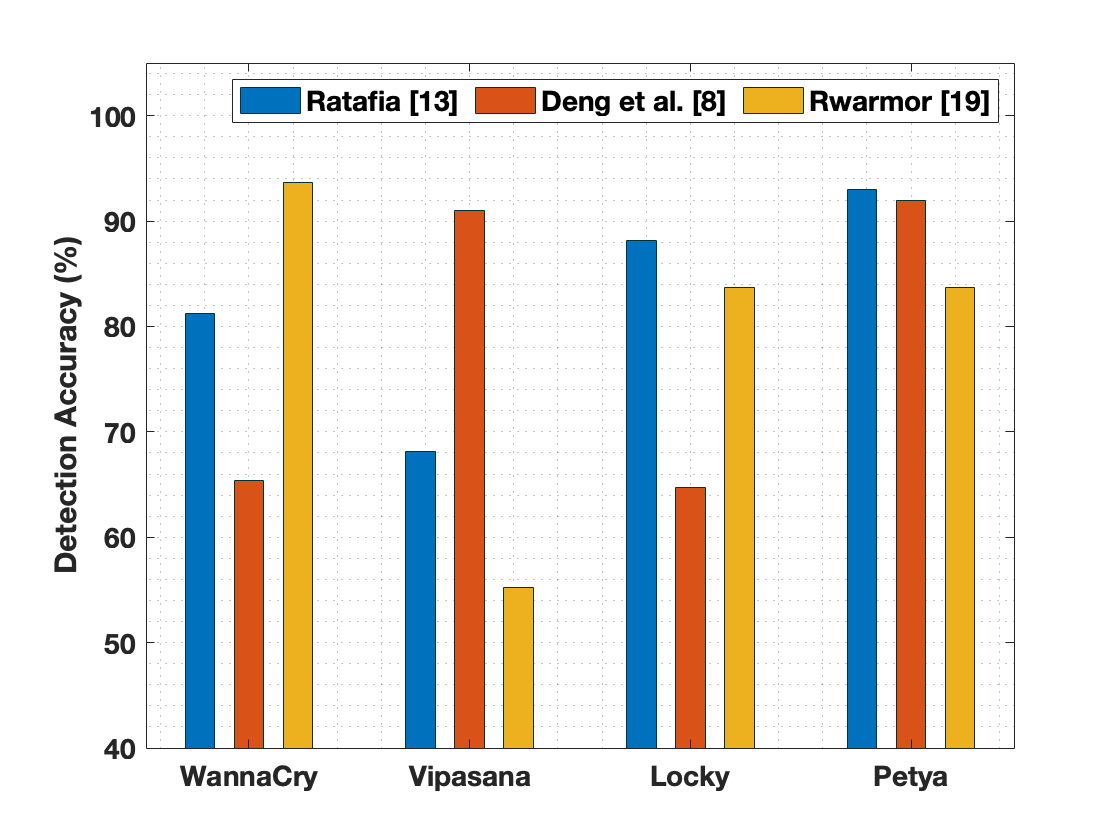}
\vspace{-0.1in}
\caption{Detection accuracy of various methods against four ransomware variants.}
\vspace{-0.1in}
\label{fig:unstable}
\end{figure}

\textbf{Dynamic analysis}, on the other hand, involves executing suspicious programs to monitor runtime behavior. For example,~\cite{alam2019ratafia} leveraged runtime performance traces to identify high-frequency encryption patterns, enabling faster detection, while~\cite{nguyen2024using} investigated a framework that tracks dynamic variations in portable executable headers. Several other studies~\cite{gulmez2024xran,herrera2023dynamic} have incorporated machine learning (ML) models to further improve detection accuracy. Despite promising results, these methods still face several challenges. First, many fail to enforce early termination of ransomware, allowing significant data encryption before detection. Second, ML-based systems often behave as black boxes without interpretable reasoning, which undermines user confidence in critical applications. Finally, most approaches are evaluated only on limited ransomware families, reducing their generalizability. As shown in Fig.~\ref{fig:unstable}, existing algorithms exhibit fluctuating performance across four representative variants—\textit{WannaCry}, \textit{Vipasana}, \textit{Locky}, and \textit{Petya}—with detection rates occasionally dropping below 70\%.

\section{Proposed Method}
\label{sec:prop}
\subsection{Overview}
To highlight the motivation for our proposed method, we summarize the key limitations of existing anti-ransomware techniques discussed in Section~\ref{sec:bgd}, as follows: 
\begin{itemize}
    \item Vulnerability to obfuscation.
    \item Lack of prioritization of early termination.
    \item Low generalizability among variants of ransomware.
    \item Lack of transparency in decision making. 
\end{itemize}

\begin{figure}[htbp] 
\centering 
\vspace{-0.2in} 
\includegraphics[width = 0.5\textwidth]{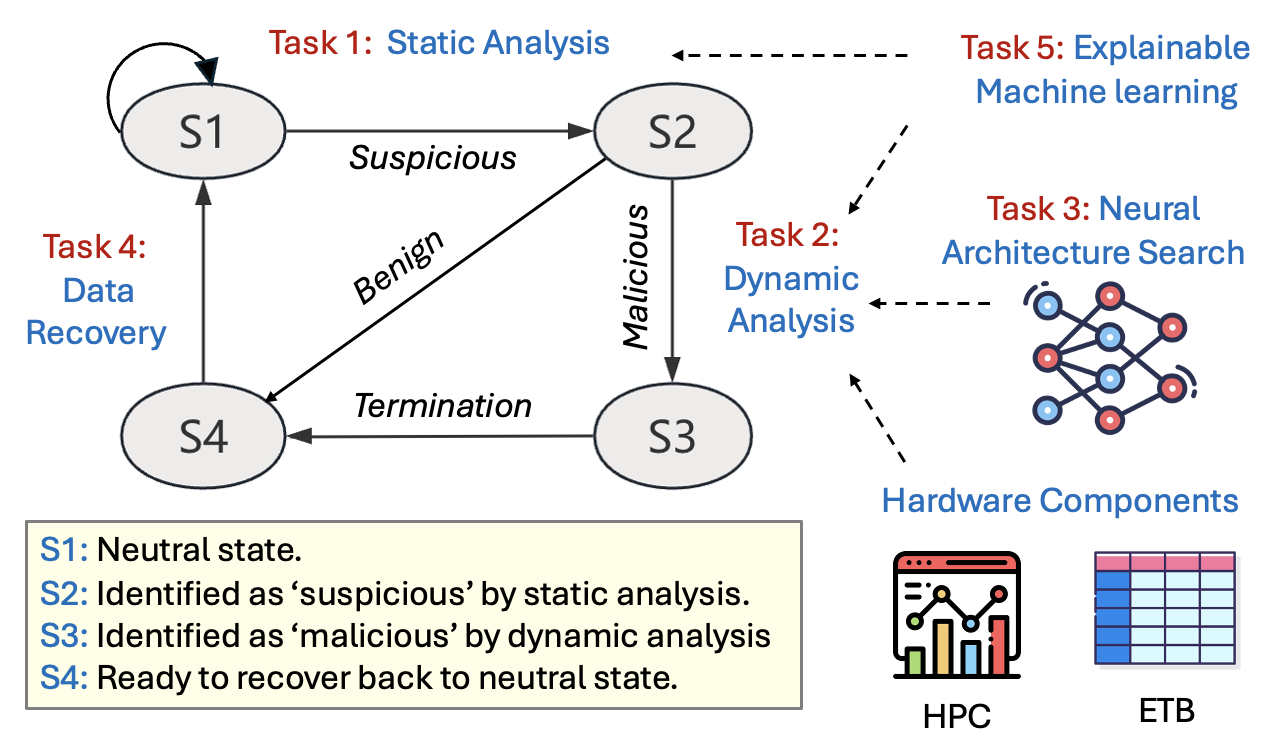} 
\vspace{-0.2in} 
\caption{State transitions of proposed framework.} 
\vspace{-0.15in} 
\label{fig:proposed} 
\end{figure}

To address these challenges, our proposed approach enables a synergistic integration of a robust static detection framework and hardware-assisted dynamic analysis for efficient detection. Figure~\ref{fig:proposed} presents an overview of the proposed workflow, modeled as a finite state machine (FSM).

In the first stage, the target program undergoes static analysis for early-stage detection. If a program is flagged as `suspicious' at this step, it is isolated to prevent further infection and passed to the dynamic analysis phase. This two-stage sequential framework offers several key advantages: it mitigates the issue of high false positive rates in conventional static analysis by labeling programs as suspicious rather than directly classifying them as malicious. Additionally, only programs flagged in the first stage proceed to runtime monitoring, which reduces unnecessary computational overhead. Next, based on the results of dynamic analysis, the program is either restored to its original state if deemed benign or, if identified as malicious, our method terminates its execution and restores encrypted files using backups.

The implementation of the workflow described above can be divided into five primary tasks:
%each of which will be detailed in the following sections: 
\begin{itemize} 
    \item A \textbf{static analysis} phase, utilizing a boosted decision tree (DT) for early detection, where adversarial training is applied to mitigate the vulnerability to obfuscation. 
    \item A \textbf{hardware-assisted dynamic analysis} phase, leveraging a long-short term memory (LSTM) model to monitor runtime behavior with features from hardware components, including hardware performance counters (HPC) and embedded trace buffer (ETB). 
    \item A \textbf{neural architecture search} (NAS) algorithm to automate the selection and optimization of machine learning models for the first two tasks. 
     \item A \textbf{data recovery} task, where backups of recently accessed files are used to restore encrypted data. 
    \item A post-detection task focuses on \textbf{explainable machine learning} analysis, which provides human-interpretable explanations of detection results in both static and dynamic analysis stage. 
\end{itemize}

\subsection{Adversarial Training based Static Analysis}
\begin{figure}[htbp]
\centering
\vspace{-0.1in}
\includegraphics[width=0.48\textwidth]{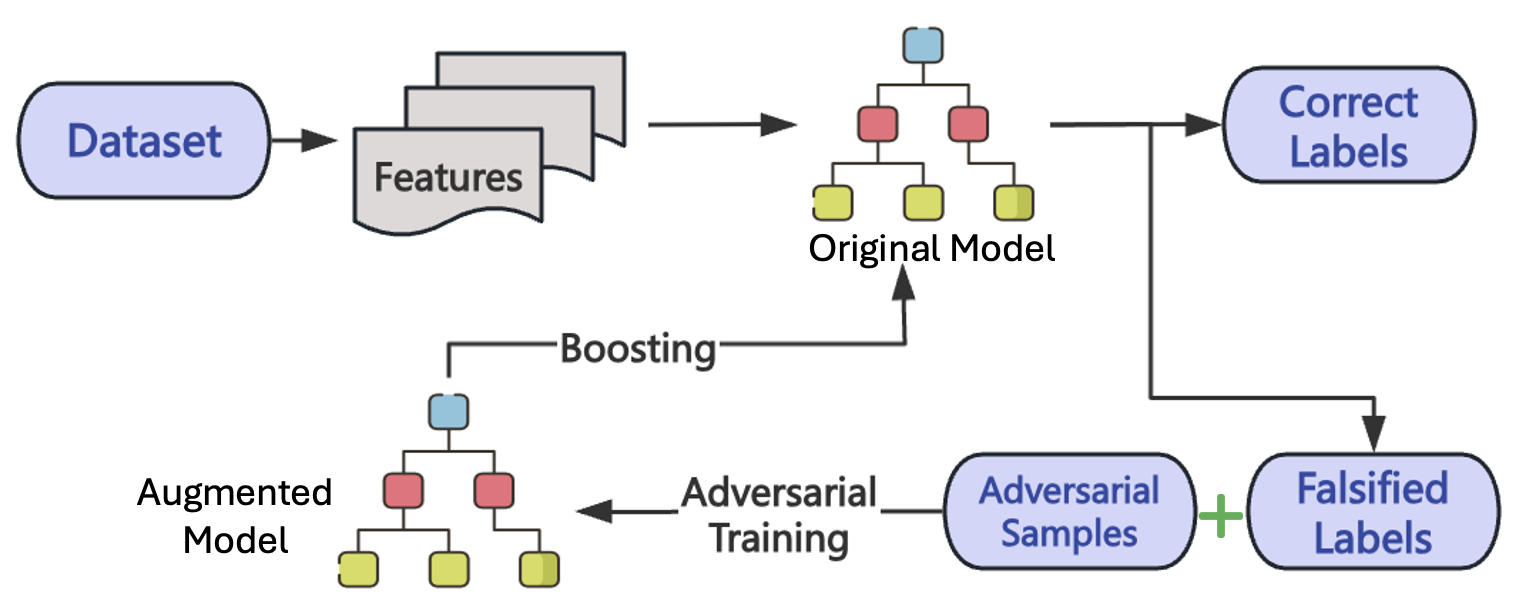}
\vspace{-0.1in}
%\caption{Workflow of adversarial training based static analysis.}
\caption{Workflow of adversarial‐training–enhanced static analysis. We synthesize {adversarial Samples} to emulate evasive/poisoned conditions. The adversarial set is combined with the clean dataset for {adversarial Training}, yielding \emph{augmented models}. A lightweight \emph{boosting} step integrates all models together, thereby improving the overall robustness.}
\vspace{-0.1in}
\label{fig:static}
\end{figure}

We employ a gradient-boosted decision tree (DT) model as the core static analysis engine. The entire workflow is shown in Figure~\ref{fig:static}. At the preprocessing stage, we extract a diverse set of static features from each program, covering API calls, opcode sequences, file metadata, Portable Executable (PE) header information, and opcode $n$-grams that capture recurring instruction patterns. These features provide complementary perspectives on program behavior, enabling the detector to capture both structural and semantic cues. For training, the dataset is split into 80\% for model training and 20\% for validation. An initial DT baseline is first constructed using the classical CART~\cite{CART} algorithm and evaluated to identify samples that are misclassified, which serve as valuable ``falsified labels'' for further refinement.

After obtaining the initial DT model, we extend the framework with an adversarial training process to reinforce its robustness against evasive ransomware. Specifically, in addition to inserting benign code fragments, which simulate padding-based obfuscation, we apply a variety of advanced obfuscation strategies:  
% \begin{enumerate}
%     \item \textbf{Opcode substitution: } replacing selected instructions with semantically equivalent alternatives;  
%     \item \textbf{Control-flow manipulation:}  reordering benign segments or injecting redundant jumps;  
%     \item \textbf{Polymorphic transformation:}  introducing randomized but functionally consistent code mutations.  
% \end{enumerate}
\begin{enumerate}
    \item \textbf{Opcode substitution:} replacing selected instructions with semantically equivalent alternatives (e.g., swapping instruction sequences that perform the same computation;
    \item \textbf{Control-flow manipulation:}  reordering benign code blocks or injecting harmless jumps/NOPs to alter control-flow structure without changing functionality;
    \item \textbf{Polymorphic transformation:}  applying lightweight syntactic mutations such as register renaming and insertion of no-op instructions to produce many syntactic variants of the same logic.  
\end{enumerate}

These transformations mimic practical obfuscation techniques employed by modern ransomware to evade static signature-based tools. Both the adversarially crafted samples and the previously misclassified instances are fed back into the training pipeline, ensuring that the model incrementally learns to recognize obfuscated ransomware while maintaining a low false positive rate.

A central challenge of adversarial training is achieving convergence under such adversarial conditions~\cite{wang2021convergence}. To overcome this, we adopt a boosting framework~\cite{ke2017lightgbm}, where a sequence of lightweight DT models $\mathcal{M} = \{\mathcal{T}_1,\mathcal{T}_2,\dots,\mathcal{T}_N\}$ is progressively trained with adaptive weights $\{w_1,w_2,\dots,w_N\}$. In each iteration, a new augmented tree $\mathcal{T}_i$ is trained specifically on the challenging and adversarial samples. The ensemble prediction is obtained via weighted voting, where higher weights are assigned to trees with stronger performance on the augmented dataset. This iterative scheme steadily enhances the detector’s resilience to obfuscation while keeping the training process stable and computationally efficient. The overall process is summarized in Algorithm~\ref{alg:alg1}.

\vspace{-0.1in}
\begin{algorithm}[htp]
%\SetAlgoLined
\SetKwInOut{Input}{Input}
\SetKwInOut{Output}{Output}
\Input{Dataset($\mathcal{D}$), epochs $(k)$}
\Output{Model $\mathcal{M}$} 
 Initialize: $\mathcal{M} = \emptyset$, $k = 0$\\
%$N = $ sizeof ($\mathcal{D}$), $M = $ sizeof ($\mathcal{F}$), $t= 0$\\
%For each $X_i \in \mathcal{D}$, $w^b_{i_(t)} = \frac{1}{N}$\\
 \Repeat{$k \geq max\_epoch$ or convergence}{
  Split $\mathcal{D}$ into training set $\mathcal{D}_t$ and validation set $\mathcal{D}_v$  \\
% $\mathcal{F}$ = extract($\mathcal{D}_t$)  \hfill\Comment{Feature extraction}\\
% Random Sample $\mathcal{D}' \subset \mathcal{D}, \mathcal{F} \subset \mathcal{F}'$ \hfill\Comment{Data Sampling}\\
 $\mathcal{T}_{(k)} = $ train($\mathcal{D}_t$) \hfill\Comment{Model training}\\
 $\mathcal{D}' \,\,=$ validate($\mathcal{T}_{(k)}$, $\mathcal{D}_v$)\hfill\Comment{Falsified labels}\\
 $\hat{\mathcal{D}}\,\,\,=$ craft($\mathcal{D}$) \hfill\Comment{Craft adversarial samples}\\
 $\mathcal{T}_{(k+1)}$ =  train($\mathcal{D}'\cup \hat{\mathcal{D}})$ \hfill\Comment{Adversarial training}\\
 $\mathcal{D} = \mathcal{D}\cup \mathcal{D}' \cup \hat{\mathcal{D}}$ \hfill\Comment{Dateset augmentation}\\ 
 $\mathcal{M} = \mathcal{M} \cup \{\mathcal{T}_{(k)}\}$\\
 \For {each $\mathcal{T}_i \in \mathcal{M}$}{
 $acc_i = 1 - |$validate($\mathcal{T}_{(k)}$)$|$/$|\mathcal{D}|$ \\
 $w_i = \frac{acc_i}{\sum\limits^N_{i=1} acc_i}$\hfill\Comment{Weight Adjustment}\\
 } 
$k = k+1$
}
Return $\mathcal{M}$\\
\caption{Adversarial training of boosting DTs}
\label{alg:alg1}
\end{algorithm}

\vspace{-0.2 in}
\subsection{Hardware-assisted dynamic analysis}
\label{sec:hardware}

Once a target program is flagged as ``suspicious'' by the static stage, it is passed to dynamic analysis for further inspection. The key challenge here is that ransomware must be identified and terminated before large amounts of files are encrypted. Relying solely on software-level monitoring often introduces unacceptable delays. To overcome this, we utilize hardware components that can stream low-level execution information in real time, thereby reducing detection latency.

\begin{figure}[htbp]
\centering
%\vspace{-0.1in}
\includegraphics[width = 0.51\textwidth]{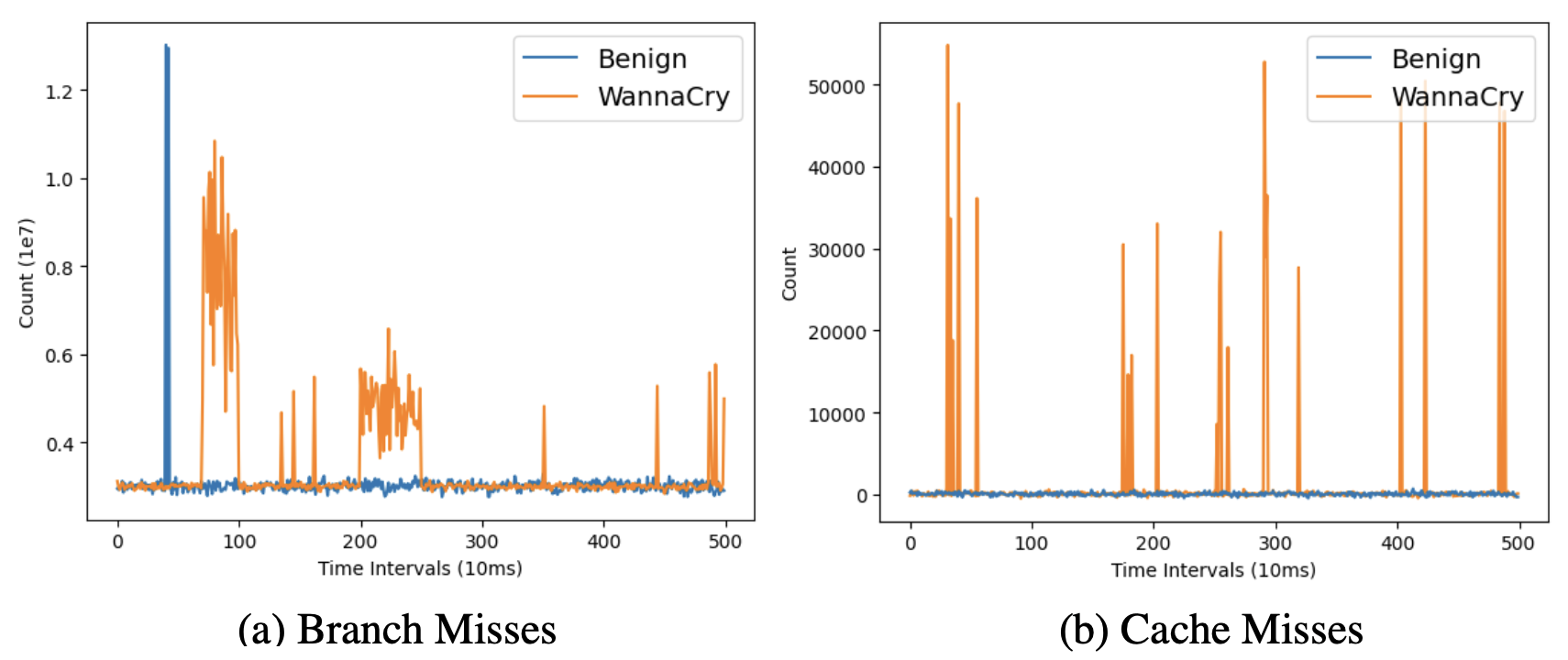}
\vspace{-0.1in}
\caption{Comparison of HPC events between the benign program and WannaCry. a) branch misses, b) cache misses.}
\label{fig:hpc}
\end{figure}

\paragraph{HPC features.}
The first hardware source we leverage is hardware performance counters (HPCs), and the potential of HPCs for ransomware detection has been explored in~\cite{alam2019ratafia}. Specifically, HPCs are built-in registers capable of tracking processor events such as instruction counts, cache references, and branch mispredictions. These metrics can be collected using standard interfaces like \textit{perf}. Importantly, many ransomware behaviors manifest distinct hardware footprints. For instance, Figure~\ref{fig:hpc} illustrates the clear gap between a benign program and ransomware (WannaCry) in terms of branch and cache misses. The frequent encryption loops in WannaCry lead to repeated branching and file access, which manifest as spikes in these HPC signals. 

\paragraph{ETB features.}
Though HPC features act as a valid way to enable realtime ransomware trace monitoring, they only provide an aggregate view of processor activity, making it unable to pinpoint activities to a particular program. To address this limitation, we also integrate embedded trace buffers (ETBs) into our framework. ETBs capture detailed, cycle-by-cycle on-chip trace for designated process, offering more precise and granular insights than HPCs. However, it comes at the cost of increased overhead, making continuous monitoring with ETBs impractical.

\paragraph{Adaptive monitoring.}
To combine the efficiency of HPCs with the precision of ETBs, we design an \textit{adaptive monitoring strategy}. Instead of following a fixed alternation schedule between lightweight and heavyweight monitors, our method adaptively determines when to escalate from HPC-based monitoring to ETB confirmation. Specifically, each program execution is divided into time windows (500\,ms in our experiments). During each window, the system performs five major steps:

\begin{enumerate}
    \item \textbf{Initial monitoring with ETBs}: In the initial window, we use ETBs for monitoring, since the program has already been deemed suspicious by the static analysis. The assumption is that it has a high likelihood of being ransomware, so ETBs are employed to detect any malicious behavior at the earliest stage.
    \item \textbf{Risk-guided monitoring with HPCs}  If the program passes the first ETB-based window without exhibiting malicious behavior, we assume it may either be benign or a ransomware using delayed infection tactics. To reduce the overhead of continuous ETB monitoring, we switch to HPC-based detection. The HPC classifier would compute the probability $p_t$ indicating how likely the current process exhibits ransomware behavior, alongside two supporting signals:
    \begin{itemize}
        \item \emph{Smoothed risk $\hat{p}_t$:} To avoid reacting to noisy fluctuations, we maintain an exponential moving average $\hat{p}_t = \alpha p_t + (1-\alpha)\hat{p}_{t-1}$. Intuitively, this acts like a ``memory'' of recent windows, stabilizing risk estimation in the same way an investor smooths stock prices to spot trends.
        \item \emph{Slope $\Delta p_t$:} We also monitor the change $\Delta p_t = p_t - p_{t-1}$. A sharp rise indicates that suspicious activity is accelerating, which is critical in ransomware detection because encryption routines often intensify abruptly once triggered.
        \item \emph{Uncertainty $u_t$:} Finally, we measure how confident the model is in its prediction. We define uncertainty as the predictive entropy $u_t = -(p_t \log p_t + (1-p_t)\log(1-p_t))$, which is highest when the model is near 50/50 and therefore unsure. In this context, high uncertainty means we should not fully trust the HPC model and should seek more detailed verification.
    \end{itemize}

    \item \textbf{Escalation:} If $\hat{p}_t$ exceeds a high threshold $T_{\text{hi}}$, or if the slope $\Delta p_t$ indicates a sudden jump, or if uncertainty $u_t$ is large, then ETB tracing is activated in the next window. This escalation means we switch from coarse monitoring to fine-grained tracing for confirmation.

    \item \textbf{ETB confirmation:} Once escalation is triggered, ETBs are enabled to capture fine-grained traces of the suspicious process. This step acts as a high-confidence verifier. If the ETB classifier confirms ransomware behavior, the process is immediately terminated. Otherwise, we consider that it indicates either the detection result of previous window is a false positive, or the ransomware applies obfuscation to circumvent detection. 
    % Either way, we perform data restoration with backup files to erase the possible infection in previous windows, and switch back to HPC-based detection phase.  
    \item \textbf{Rollback and recovery:} Either the ransomware activity was confirmed or a conflict between ETB and HPC happened, we perform data restoration with backup files to erase the possible infection in previous windows (details in Section~\ref{sec:recover}). This ensures that any partial encryption is reversed to mitigate the ransomware damage.
     \end{enumerate}

\begin{figure}[htbp]
\centering
\includegraphics[width=0.5\textwidth]{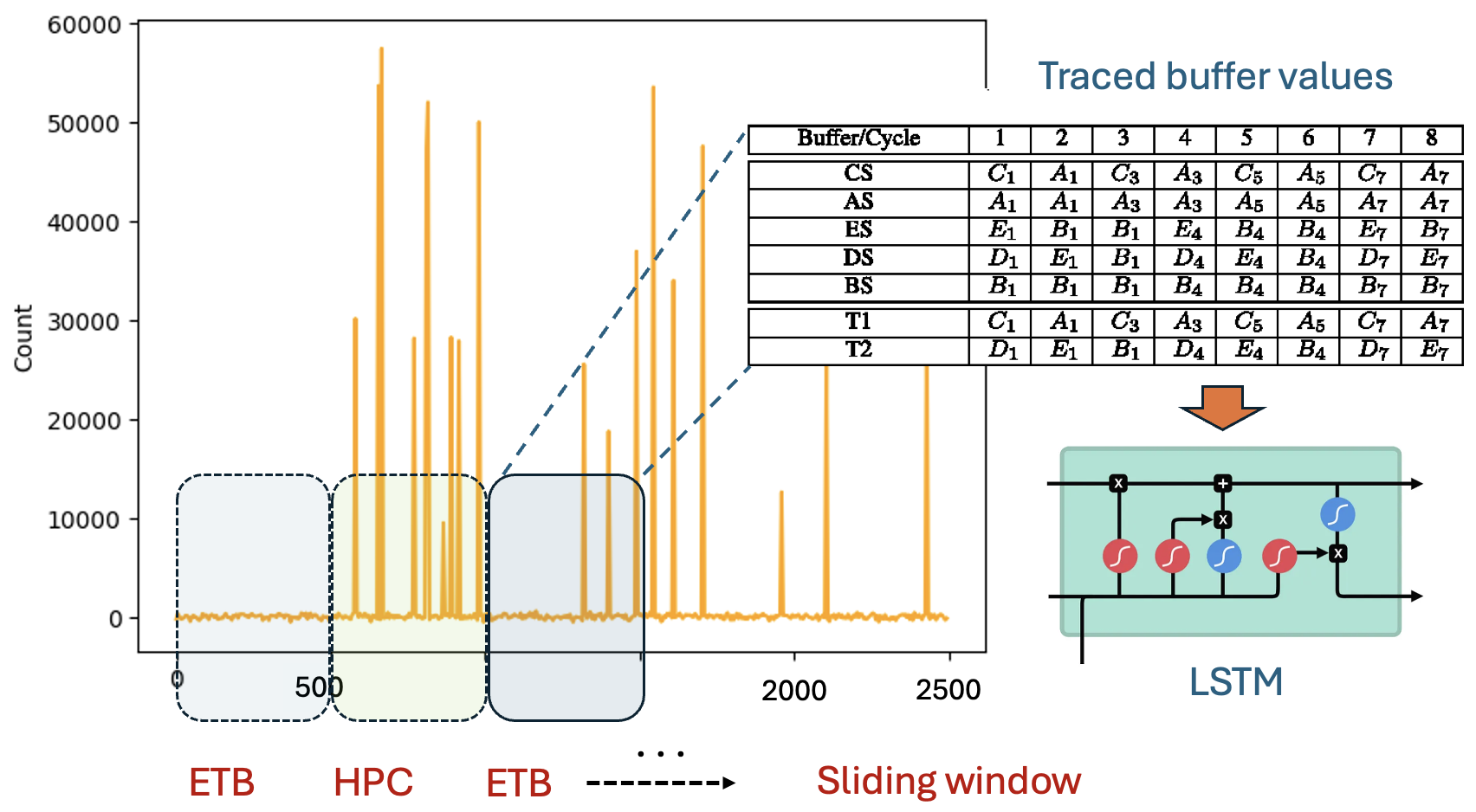}
\caption{Adaptive monitoring workflow combining HPC and ETB. An LSTM model is applied to classify the sequential hardware traces.}
%\vspace{-0.1 in}
\label{fig:cmb}
\end{figure}

The workflow is shown in Figure~\ref{fig:cmb}. We select 500ms as the window size to efficiently monitor ransomware activities while keeping resource usage low. In order to handle time-sequential data dumped from HPCs/ETBs, we apply long-short term memory (LSTM) as our machine learning model for classification. The details of the model structure and hyperparameter tuning will be discussed in the next section.
% Figure

\subsection{Neural Architecture Search}
In this section, we elaborate on how we determine the structure and parameters for the applied LSTM model. As illustrated in Section~\ref{sec:bgd}, another key challenge in ML-based ransomware detection is the limited generalizability of models across different variants. We attribute this challenge to the fact that in most existing approaches, model architectures were manually selected based on prior knowledge, which lacks the flexibility to adapt to unseen variants. A straightforward approach could be to exploit a large-scale model. However, such models introduce high latency and are more prone to overfitting~\cite{ying2019overview}, which is undesirable in a latency-sensitive application such as ransomware detection.

To address this issue, we employ Neural Architecture Search (NAS) to automatically discover optimal model architectures tailored for our task. NAS explores a large space of potential network configurations, evaluating candidate architectures with different layers and connections, and gradually converges to structures that achieve high task-specific performance~\cite{elsken2019neural}. 
%The general workflow of NAS is depicted in Figure~\ref{fig:elsken}. 

% \begin{figure}[htbp]
% \centering
% \vspace{-0.1 in}
% \includegraphics[width = 0.5\textwidth]{figures/NAS.png}
% \vspace{-0.1 in}
% \caption{General workflow of Neural Architecture Search (NAS). (Image credit: Elsken, et al. 2019~\cite{elsken2019neural}).}
% \vspace{-0.1 in}
% \label{fig:elsken}
% \end{figure}

Our framework employs a one-shot NAS paradigm for better efficiency. Instead of training each candidate architecture individually, we first construct a large, over-parameterized \textit{supernet} that contains many possible sub-architectures. Then we progressively eliminate redundant components while retaining helpful ones, resulting in a final optimized model architecture that is both efficient and effective. The detailed workflow proceeds in three stages: 

% \begin{itemize}
%     \item \textbf{Supernet training.} The supernet encompasses diverse architectural choices, including LSTM cells, convolutional layers, fully connected layers, max pooling, and activation functions. Each subnet is sampled and trained as part of the supernet optimization, allowing knowledge sharing across candidates.
%     \item \textbf{Pruning with multi-objective criteria.} Once the supernet has converged, candidate subnets are pruned by progressively disabling redundant nodes and connections. Unlike conventional approaches that focus purely on accuracy, we incorporate both detection accuracy and inference latency into the evaluation as ransomware detection is a latency-sensitive task. In this case, each candidate is assigned a composite loss:
%     \[
%     \mathcal{L}_{\text{NAS}} = \lambda \cdot (1-\text{Acc}) \;+\; (1-\lambda)\cdot \frac{\text{Latency}}{\text{Latency}_{\max}},
%     \]
%     where $\text{Acc}$ denotes detection accuracy, $\text{Latency}$ is the average runtime per detection window, $\text{Latency}_{\max}$ normalizes the scale, and $\lambda \in [0,1]$ controls the accuracy/latency trade-off. Larger $\lambda$ values bias the search toward highly accurate but deeper models; Instead, smaller $\lambda$ favors lighter models with faster response time.
% \end{itemize}

\begin{enumerate}
    \item \textbf{Supernet training.} We first construct an over-parameterized supernet representing the entire search space, including LSTM units, convolutional and fully connected layers, pooling operations, and activation functions. The supernet is trained end-to-end for the detection task, so that all components inherit useful parameters.

    \item \textbf{Subnet evaluation.} After supernet training, we sample a population of candidate subnets $a \in \mathcal{A}$. For each $a$, we measure its accuracy $\mathrm{Acc}(a)$ and latency $\mathrm{Lat}(a)$, and compute the composite loss as
\[
\mathcal{L}_{\mathrm{NAS}}(a) = \lambda\,(1-\mathrm{Acc}(a)) + (1-\lambda)\,\frac{\mathrm{Lat}(a)}{\max_{j}\mathrm{Lat}(a_j)}.
\]

Next, to relate these losses to architectural choices, we encode each sampled architecture as a binary vector over its $m$ selected components, which gives us a matrix $X \!\in\! \{0,1\}^{P \times m}$ for $P$ samples. We then approximate the observed losses
\[
L = \big[\mathcal{L}_{\mathrm{NAS}}(a_1), \dots, \mathcal{L}_{\mathrm{NAS}}(a_P)\big]^\top
\]
as a linear combination of the selected components, with coefficient vector $\beta = \{\beta_1,\beta_2,\dots, \beta_m\}$ representing the contribution of each component. To encourage compactness and prevent overfitting, we further impose an $\ell_1$ penalty, leading to the optimization problem:
\[
\beta \;=\; \arg\min_{\beta}\;\Big(\|L - X\beta\|_2^2 + \gamma \|\beta\|_1\Big),
\]
where $\gamma$ controls the sparsity of the solution. By solving this optimization problem, we obtain $\beta$ as the evaluation of every component involved in this iteration, which helps us to refine the architecture.

\vspace{0.1 in}
\item \textbf{Pruning.} Components with large positive coefficients, indicating their inclusion to increase the composite loss, are pruned from the search space using a threshold $\tau$ set relative to the distribution of coefficients. The supernet is then retrained on the reduced space, and this process repeats until convergence.
\end{enumerate}

\begin{figure}[htbp]
\centering
%\vspace{-0.05in}
\includegraphics[width = 0.48\textwidth]{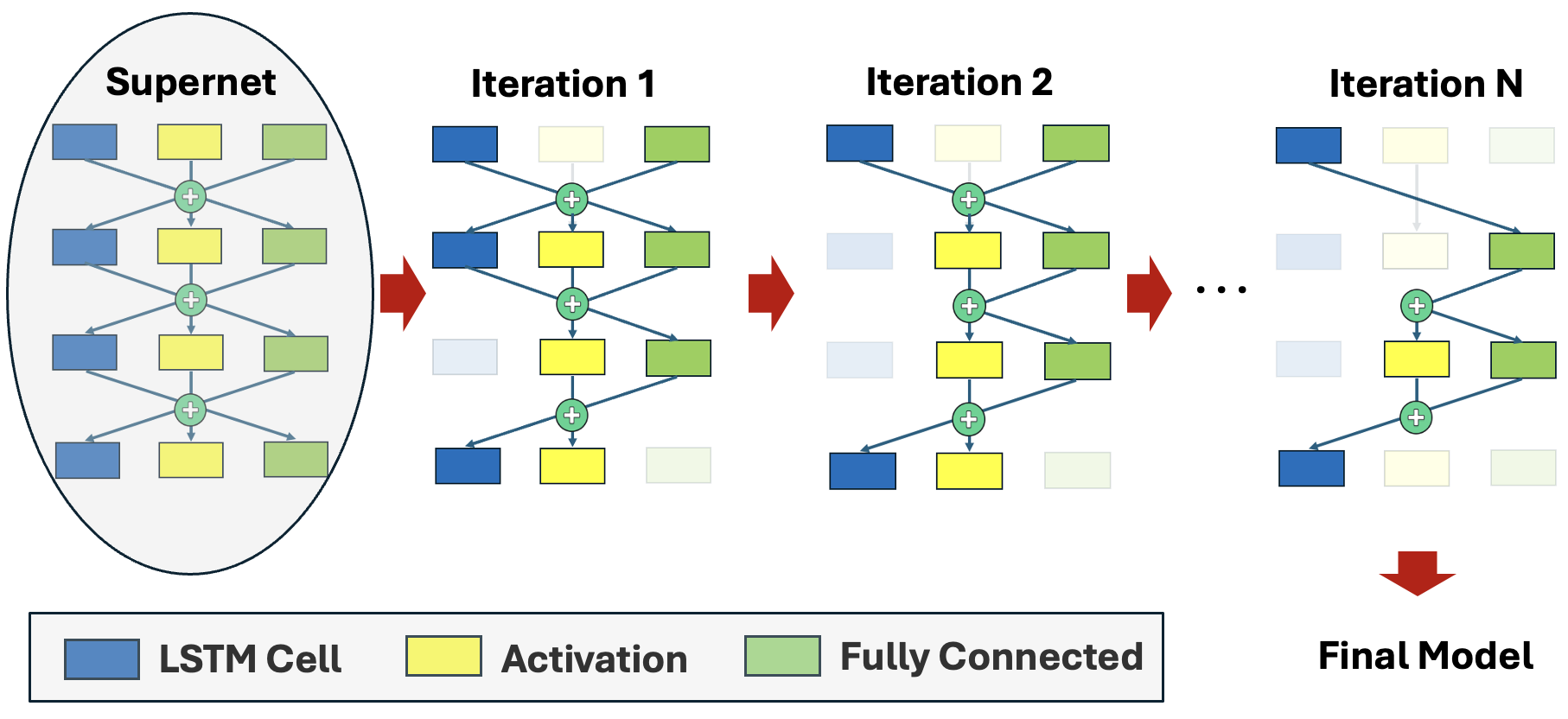}
%\vspace{-0.1in}
\caption{The proposed NAS process: Starting from an over-parameterized supernet, candidate sub-architectures are sampled and evaluated with the composite loss. Then the components (cells) associated with higher loss contributions are pruned (shown faded), the process continues until a compact final model is obtained.
}
\vspace{-0.1in}
\label{fig:oneshot}
\end{figure}

%This iterative workflow progressively eliminates redundant components while retaining helpful ones. 
% The architecture becomes progressively more streamlined, resulting in a final model that is both efficient and effective for ransomware detection.
\vspace{0.1 in}
An illustration of this NAS pruning process is shown in Figure~\ref{fig:oneshot}. We start with a dense supernet, where redundant nodes are progressively removed, resulting in a streamlined architecture optimized for both effectiveness and efficiency. The entire workflow incorporating all three major steps is presented in Algorithm~\ref{alg:alg2}.

\begin{algorithm}[htp]
\SetKwInOut{Input}{Input}
\SetKwInOut{Output}{Output}
\Input{Dataset $\mathcal{D}$, rounds $K$, trade-off $\lambda$, regularization $\gamma$, threshold $\tau$, population $P$}
\Output{Optimized architecture $\mathcal{M}$}
Initialize supernet $\mathcal{S}$; $r=0$ \\
\Repeat{$r \geq K$ or convergence}{
   Train $\mathcal{S}$ on $\mathcal{D}$ (weight sharing) \\
   Sample $P$ candidate architectures $\{a_1,\dots,a_P\}$ \\
   For each $a_i$: compute $\mathrm{Acc}(a_i)$, $\mathrm{Lat}(a_i)$ \\
   $\tilde{L}(a_i)=\mathrm{Lat}(a_i)/\max_j \mathrm{Lat}(a_j)$ \\
   $L_i=\lambda(1-\mathrm{Acc}(a_i))+(1-\lambda)\tilde{L}(a_i)$ \\
   Build design matrix $X\in\{0,1\}^{P\times m}$ \\
   $\beta=\arg\min_\beta \|L-X\beta\|_2^2+\gamma\|\beta\|_1$ \\
   Prune components with $\beta_i \geq \tau$ \\
   Update $\mathcal{S}$; $r=r+1$ \\
}
Select $\mathcal{M}$ with minimum $L_i$ \\
\Return $\mathcal{M}$ \\
\caption{NAS with composite loss and sparse regression pruning}
\label{alg:alg2}
\end{algorithm}

\vspace{-0.15 in}
\subsection{Data Recovery}
\label{sec:recover}

Although our proposed model allows for real-time detection and significantly reduces latency, some files may still inevitably get encrypted by ransomware before the malicious activity is flagged. To mitigate this impact, we extend our framework with a dynamic backup and recovery framework that runs in parallel with the monitoring process. This module complements detection by providing a rapid rollback capability, thereby completing the end-to-end defense cycle. The entire framework consists of three major modules.

\paragraph{Lightweight snapshotting.}  
The first module adopts a rolling snapshot approach to keep track of file changes. During program execution, we run a shell script using \texttt{rsync} command to save lightweight copies of the modified files inside the focused directory. Notice that instead of maintaining full backups of entire directories, this script executes at one-minute intervals and captures only those files that have been accessed or modified during the latest window. 
%This incremental strategy minimizes storage overhead while ensuring effective file protection. 

\paragraph{Integration with detection.}  
The snapshotting process runs in parallel with the dynamic analysis module (Section~\ref{sec:hardware}). In each time window, a temporary backup is created using lightweight snapshotting for files that were accessed. If the LSTM detector concludes that the program behavior is benign, the snapshot is discarded to conserve resources. Otherwise, as the detector raises an alert indicating ransomware activity, the program is terminated and the snapshot is used to roll back any affected files. This parallel design keeps backups continuously available, enabling rapid recovery with minimal latency.

\paragraph{File event logging.}  
To make the framework more practical for real-world use, we incorporate an \texttt{inotify}-based logger that continuously records file system events (e.g., \textit{open}, \textit{modify}, \textit{create}, \textit{delete}) within the monitored directories. This logging helps confirm that snapshots and recoveries happen as intended, and it also keeps a detailed record of file activity. Such records allow administrators to verify what happened to each file and restore the sequence of actions if any incident happened.

% To support this process and enable forensic analysis, we also employ an \texttt{inotify}-based logger that records file system events (e.g., \textit{open}, \textit{modify}, \textit{create}, \textit{delete}) within the monitored directories. This logging not only verifies that snapshots are triggered correctly but also provides a per-file audit trail, allowing administrators to reconstruct the sequence of file operations during an incident.

% \begin{figure}[htbp]
% \centering
% %\vspace{-0.05in}
% \includegraphics[width = 0.35\textwidth]{figures/image002.png}
% %\vspace{-0.1in}
% \caption{The proposed NAS process.}
% %\vspace{-0.1in}
% \label{fig:inotify}
% \end{figure}

% The recovery mechanism thus augments detection with three unique advantages:  

%     \item \textbf{Efficiency:} only modified files are snapshotted, minimizing storage and runtime costs.  
%     \item \textbf{Responsiveness:} snapshots can be restored immediately once ransomware is flagged, limiting the damage window to less than one minute.  
%     \item \textbf{Auditability:} per-file event logs provide forensic visibility, enabling administrators to trace attack behavior and verify recovery actions.  
% \end{itemize}
% Together, these features ensure that even if ransomware partially executes before detection, its impact on user data is contained and recoverable.

In summary, the proposed data recovery strategy complements our detection pipeline, it minimizes storage overhead while ensuring effective file protection. 

\subsection{Explainable Machine Learning}
\label{sec:xai}

% Another major limitation of existing methods is the lack of interpretability. Due to the black-box nature of ML models, no additional information apart from the detection result is available. On the one hand, system administrators have no guidance on how to respond to misclassifications. On the other hand, even when predictions are correct, a binary label alone is often insufficient since administrators may need additional information. To address this, we integrate explainable machine learning into both the static and dynamic components of our framework, enabling not only accurate classification but also actionable insights for follow-up tasks such as feature refinement and temporal localization.

Another major limitation of existing methods is the lack of interpretability. Due to the black-box nature of ML models, no additional information beyond the detection result is available. In such cases, users have no guidance on how to respond to misclassifications. Moreover, a binary label alone may be insufficient for follow-up tasks such as ransomware localization. To address this limitation, we integrate explainable machine learning into both static and dynamic analysis.

% For static analysis, we adopt SHAP for interpretability. SHAP originates from cooperative game theory, where the goal is to measure the marginal contribution of each player to the overall outcome. Formally, given a feature set $N$, a subset $S \subseteq N$ excluding feature $i$, and model function $f(\cdot)$, the SHAP value of feature $i$ is defined as
% \begin{equation}
% \phi_i = \sum\limits_{S \subseteq N \setminus \{i\}} \frac{|S|!(|N|-|S|-1)!}{|N|!}\,\big[f(S \cup \{i\}) - f(S)\big].
% \end{equation}
% In our context, SHAP values $\phi_i$ represent the average contribution of a feature (e.g., a specific API call or opcode sequence) to the classification result. In this way, we could use SHAP both for misclassification handling and outcome interpretation. On the one hand, we could report the top-$k$ features with the highest $|\phi_i|$ for each flagged sample, explaining the pattern (API-based, event-triggered) of ransomware behavior. On the other hand, features with persistently low contributions can be pruned or down-weighted in subsequent training iterations, reducing noise and improving generalization for our boosted detection model.

For static analysis, we adopt SHAP~\cite{lundberg2017unified} to enhance interpretability. SHAP originates from cooperative game theory, where the goal is to quantify the marginal contribution of each player to the overall outcome. Formally, given a feature set $N$, a subset $S \subseteq N$ excluding feature $i$, and model function $f(\cdot)$, the SHAP value of feature $i$ is defined as
\begin{equation}
\phi_i = \sum\limits_{S \subseteq N \setminus \{i\}} \frac{|S|!(|N|-|S|-1)!}{|N|!}\,\big[f(S \cup \{i\}) - f(S)\big].
\end{equation}
In our context, $\phi_i$ represents the average contribution of a static feature (e.g., an API call or opcode sequence) to the classification result. We employ SHAP for both misclassification handling and outcome interpretation. For each flagged sample, we report the top-$k$ features with the highest $|\phi_i|$, which helps explain underlying ransomware patterns such as API-based triggers. In addition, features with consistently low contributions during trainning phase would be pruned, reducing noise and improving the performance of the detection model.

As for dynamic analysis, the model architecture is no longer fixed due to the utilization of NAS, and the input consists of sequential HPC/ETB traces with variable length. Therefore, we require a model-agnostic method that is also compatible with sequential data. We employ LIME~\cite{ribeiro2016lime}, which addresses this by fitting a local linear regression surrogate to approximate the detector’s decision based on input traces. Formally, for an input (hardware traces from HPC/ETB) $x = [x_1, x_2, \dots, x_T]$ within $T$ time slots, the surrogate prediction is given by
\begin{equation}
\hat{y} = w_0 + \sum_{t=1}^{T} w_t x_t,
\end{equation}
where $w_t$ represents the contribution of time slot $t$. In this way, the absolute values of $w_t$ actually indicate the relative importance of slots, allowing us to rank them and interpret the results in the context of ransomware detection. For example, clusters of important slots suggest concentrated malicious activity within specific intervals, while periodically separated slots may reflect recurring actions such as client–server communication. As a result, our detection model could offer localized insights into the trace patterns beyond the detection result, which will be demonstrated in Section~\ref{sec:xaiexp}.

% we treat each time slot (or buffer cycle) as an interpretable unit. LIME perturbs subsets of slots, fits a sparse linear regression surrogate, and assigns coefficients to each slot. Once coefficients are obtained, they are ranked to identify critical windows. Large positive coefficients indicate slots strongly associated with ransomware behavior. Interpreting these rankings in context provides meaningful insights. For example, clusters of adjacent important slots suggest that malicious activity is concentrated in a specific time window. Periodically separated important slots may reflect repeated malicious actions, such as client–server communication where stolen data is transmitted at fixed intervals. At a finer granularity, the same process applied row-by-row allows us to identify which signals within a cycle are most likely linked to malicious operations, leading to malware localization.

% Figure~\ref{fig:xai} illustrates typical outputs, where static SHAP values highlight key features and dynamic LIME weights point to the cycles most responsible for a suspicious classification.

\section{Experimental Evaluation}
\label{sec:exp}

In this section, we demonstrate the effectiveness of our proposed framework. We first outline the experimental setup. Next, we present the performance of the proposed framework.

\subsection{Experiment Setup}

% \noindent{\underline{\bf Benchmark}}: We conducted performance evaluations on a standard host machine running Linux. Four ransomware variants \textit{WannaCry}, \textit{Vipasana}, \textit{Locky}, and \textit{Petya} were selected as benchmarks, and the SPEC integer benchmark~\cite{SPEC} was used to present benign programs. We also insert non-malicious instructions into the ransomware code as obfuscation to simulate adversarial samples. During program execution, hardware performance counter (HPC) values were collected using the perf tool at a 500ms sampling interval, and embedded trace buffer (ETB) values were retrieved via the UART port. To simulate adversarial ransomware samples, we inserted non-functional instructions into the ransomware code to create obfuscated versions.

% \subsection{Dataset}
% \label{sec:dataset}
\noindent{\underline{\bf Benchmark}}:
We evaluate our framework using the publicly available \emph{Ransomware Combined Structural Feature Dataset}, which contains structural features extracted from executable binaries collected and preprocessed.
The dataset comprises a total of 2,675 samples, partitioned into a training/validation split of 2,157 samples (80\%) and a test set of 518 samples (20\%). The training portion comprises 1,023 ransomware samples spanning 25 families and 1,134 goodware samples, while the test portion contains 385 ransomware samples and 133 goodware samples. The dataset provides multiple feature sets (PE headers, imported DLLs, API calls, section entropy, and opcode $n$-grams), as well as explicit train/test splits that we adopt in our experiments.

The original dataset contains a large number of distinct families (40+ family IDs available across versions). To keep the evaluation results concise, we organize the dataset using a two-tier scheme:

\begin{enumerate}
  \item \textbf{Category grouping} (main evaluation). We group related families into four high-level categories: \emph{Crypto-Ransomware}, \emph{Locker Ransomware}, \emph{Scareware/FakeAV}, and \emph{Miscellaneous/Hybrid}. For the main accuracy and latency tables, we report aggregated metrics per category (mean and standard deviation across families within the same category).
  \item \textbf{Representative variants} (detailed studies). To keep the experimental evaluation compact while still enabling a fine-grained analysis, we select representative sample variants per category for detailed case studies. These selected representative families (i) are widely studied in the literature and (ii) reflect diverse behavioral patterns (for example, \textit{WannaCry} (Crypto), \textit{Reveton} (Locker), \textit{FakeAV} (Scareware), and \textit{Phobos/Stop} (Hybrid)).

  % These representative variants are used for per-family figures (e.g., t-SNE/UMAP feature plots), cross-family holdout tests, and visual explainability examples.
\end{enumerate}

%Table~\ref{tab:dataset_stats} summarizes the dataset organization.
\begin{table}[htp]
  \centering
  \caption{Ransomware categories and representative families.}
  \label{tab:dataset_stats}
  \begin{tabular}{|l|l|}
    \hline
    \textbf{Category} & \textbf{Example Families} \\
    \hline
    Crypto-Ransomware & \textit{WannaCry, Locky, Ryuk, Maze, Petya} \\
    \hline
    Locker Ransomware & \textit{Reveton, WinLock} \\
    \hline
    Scareware/FakeAV  & \textit{FakeAV, Rogue AV variants} \\
    \hline
    Misc./Hybrid      & \textit{DoppelPaymer, BlackCat, Phobos} \\
    \hline
  \end{tabular}
\end{table}

To further evaluate robustness against obfuscation, we extend the current dataset with augmented adversarial samples. For example, one strategy we applied is to inject non-functional and benign instructions into ransomware binaries to produce obfuscated variants (opcode substitution, NOP insertion, register renaming, and lightweight polymorphic edits). These adversarially-obfuscated samples are mixed into the training and test splits according to the protocol described in Section~\ref{sec:ablation_adv}.

% As for the hardware-based data collection, we collect hardware performance counter (HPC) values using \texttt{perf} at a sampling interval of 500\,ms, and retrieve embedded trace buffer (ETB) traces via UART during program execution. These hardware features were utilized for real-time dynamic analysis as illustrated in Section~\ref{sec:hardware}.
\vspace{0.1 in}
\noindent{\underline{\bf Implementation Packet}}:
% The machine learning framework was implemented on a host Linux machine equipped with an Intel i9 3.70GHz CPU, 64 GB RAM, and an RTX 3080 256-bit GPU. We used PyTorch as the machine learning library. As for model training, the decision tree-based static analysis model was trained over 200 epochs, while the LSTM model consumed 500 epochs. The learning rate for the LSTM model was adjusted from 0.05 to 0.1 after 200 epochs to accelerate convergence.  Cross-validation was applied to further reduce bias in evaluation. Additionally, a dropout rate of 0.5 was used for both models to prevent overfitting.
The proposed framework was implemented and evaluated on a Linux host equipped with an Intel Core i9-10900K processor (3.70~GHz), 64~GB RAM, and an NVIDIA RTX~3080 GPU.
The operating system was Ubuntu~22.04~LTS with Linux kernel~6.16, and hardware performance counter (HPC) events were collected using the \texttt{perf} interface.
For runtime monitoring, HPC features were collected at a fixed sampling interval of 500~ms.
To improve process-level attribution and reduce cross-process interference, the monitored program was pinned to a dedicated CPU core using CPU affinity, and per-window counter deltas were aggregated into time-series feature vectors for the LSTM classifier.
The fine-grained execution traces were collected using Intel Processor Trace (PT) and parsed into sequential trace representations aligned with the same 500~ms monitoring windows.
Trace collection was enabled selectively by the adaptive monitoring policy described in Section~\ref{sec:hardware} to limit runtime overhead.

As for the machine learning components, the entire model framework was implemented in PyTorch. The static classifier was trained for 200 epochs, and the LSTM model for dynamic analysis was trained for 500 epochs, using an initial learning rate of 0.05, which was increased to 0.1 after 200 epochs. A dropout rate of 0.5 was applied to both models, and cross-validation was used to reduce evaluation bias.

\vspace{0.1in}
\noindent{\underline{\bf Approaches}}:
To enable fair comparison, we compare the performance and efficiency of the following four approaches. 
\begin{itemize}
\item {\bf DRL:} A deep reinforcement learning based static analysis method\cite{deng2024ransomware}.
\item {\bf Ratafia:} A dynamic analysis method using HPC values informed auto-encoders~\cite{alam2019ratafia}.
\item {\bf Rwarmor:} A hybrid detection framework using static-informed dynamic analysis~\cite{ayub2024rwarmor}.
\item {\bf Proposed:} The proposed hardware-assisted multistage detection method. 
\end{itemize}

%\vspace{0.1in}

\noindent{\underline{\bf Evaluation}}:
We consider the \textit{Accuracy, Recall, Precision, and F-1} Scores as the metrics for evaluation.

\begin{table*}[htp]
\centering
\caption{Detection performance across ransomware categories.}
\label{tab:results}
\begin{tabular}{|c|cccc|cccc|cccc|ccccc|}
\hline
\textbf{Category}    & \multicolumn{4}{c|}{DRL}                                                                                                 & \multicolumn{4}{c|}{Ratafia}                                                                                             & \multicolumn{4}{c|}{Rwarmor}                                                                                             & \multicolumn{5}{c|}{Proposed}                                                                                                                                       \\ \hline
                     & \multicolumn{1}{c|}{\textbf{Acc}} & \multicolumn{1}{c|}{\textbf{Prec}} & \multicolumn{1}{c|}{\textbf{Rec}} & \textbf{F-1} & \multicolumn{1}{c|}{\textbf{Acc}} & \multicolumn{1}{c|}{\textbf{Prec}} & \multicolumn{1}{c|}{\textbf{Rec}} & \textbf{F-1} & \multicolumn{1}{c|}{\textbf{Acc}} & \multicolumn{1}{c|}{\textbf{Prec}} & \multicolumn{1}{c|}{\textbf{Rec}} & \textbf{F-1} & \multicolumn{1}{c|}{\textbf{Acc}} & \multicolumn{1}{c|}{\textbf{Prec}} & \multicolumn{1}{c|}{\textbf{Rec}} & \multicolumn{1}{c|}{\textbf{F-1}} & \textbf{Improv(\%)} \\ \hline
Crypto               & \multicolumn{1}{c|}{85.2} & \multicolumn{1}{c|}{82.1} & \multicolumn{1}{c|}{86.4} & 0.87 & \multicolumn{1}{c|}{88.9} & \multicolumn{1}{c|}{87.5} & \multicolumn{1}{c|}{89.2} & 0.92 & \multicolumn{1}{c|}{81.3} & \multicolumn{1}{c|}{78.6} & \multicolumn{1}{c|}{82.5} & 0.88 & \multicolumn{1}{c|}{93.8} & \multicolumn{1}{c|}{94.5} & \multicolumn{1}{c|}{93.2} & \multicolumn{1}{c|}{0.96} & 8.7  \\ \hline
Locker               & \multicolumn{1}{c|}{76.5} & \multicolumn{1}{c|}{74.2} & \multicolumn{1}{c|}{77.8} & 0.82 & \multicolumn{1}{c|}{80.2} & \multicolumn{1}{c|}{81.0} & \multicolumn{1}{c|}{79.6} & 0.88 & \multicolumn{1}{c|}{74.9} & \multicolumn{1}{c|}{73.4} & \multicolumn{1}{c|}{75.5} & 0.84 & \multicolumn{1}{c|}{91.7} & \multicolumn{1}{c|}{92.1} & \multicolumn{1}{c|}{91.4} & \multicolumn{1}{c|}{0.95} & 11.5 \\ \hline
Scareware     & \multicolumn{1}{c|}{72.1} & \multicolumn{1}{c|}{70.8} & \multicolumn{1}{c|}{73.2} & 0.80 & \multicolumn{1}{c|}{79.4} & \multicolumn{1}{c|}{80.2} & \multicolumn{1}{c|}{78.6} & 0.86 & \multicolumn{1}{c|}{70.8} & \multicolumn{1}{c|}{71.0} & \multicolumn{1}{c|}{70.6} & 0.82 & \multicolumn{1}{c|}{89.9} & \multicolumn{1}{c|}{90.3} & \multicolumn{1}{c|}{89.6} & \multicolumn{1}{c|}{0.94} & 12.3 \\ \hline
Hybrid/Misc          & \multicolumn{1}{c|}{79.6} & \multicolumn{1}{c|}{77.1} & \multicolumn{1}{c|}{80.5} & 0.84 & \multicolumn{1}{c|}{84.3} & \multicolumn{1}{c|}{85.0} & \multicolumn{1}{c|}{83.7} & 0.90 & \multicolumn{1}{c|}{77.5} & \multicolumn{1}{c|}{76.9} & \multicolumn{1}{c|}{77.8} & 0.86 & \multicolumn{1}{c|}{92.3} & \multicolumn{1}{c|}{92.9} & \multicolumn{1}{c|}{91.8} & \multicolumn{1}{c|}{0.95} & 10.9 \\ \hline
\textbf{Average}     & \multicolumn{1}{c|}{78.4} & \multicolumn{1}{c|}{76.0} & \multicolumn{1}{c|}{79.5} & 0.83 & \multicolumn{1}{c|}{83.2} & \multicolumn{1}{c|}{83.4} & \multicolumn{1}{c|}{82.8} & 0.89 & \multicolumn{1}{c|}{76.1} & \multicolumn{1}{c|}{75.0} & \multicolumn{1}{c|}{76.6} & 0.85 & \multicolumn{1}{c|}{91.9} & \multicolumn{1}{c|}{92.5} & \multicolumn{1}{c|}{91.5} & \multicolumn{1}{c|}{0.95} & 10.8 \\ \hline
\end{tabular}
\end{table*}

\subsection{Case Study: Detection Accuracy}
\label{sec:results}

% Table~\ref{tab:results} presents the performance comparison between the proposed detection method and existing approaches across various ransomware benchmarks. We evaluated all four methods using accuracy (Acc), false positive rate (FPR), false negative rate (FNR), and F1 score (F1). Each row represents a specific ransomware benchmark, with average values at the end. As we can see, DRL achieves an average accuracy of 78.5\%, while Ratafia is at 83.9\%. Both methods fall short compared to our proposed approach. This is due to inherent limitations in conventional static and dynamic analysis techniques. DRL, being a static analysis method, suffers from a high false positive rate, resulting in the highest FPR among the four methods. Ratafia as a dynamic analysis approach, relies solely on HPC events without assistance from ETB, leading to a relatively high FNR. For instance, it only achieves 73.1\% accuracy on the Vipasana ransomware. Rwarmor faces an even greater challenge, managing only 58.2\% accuracy on Vipasana due to its unique behavior of encrypting files offline, without communication with control servers, making its dynamic behavior distinct from other ransomware. In contrast, our proposed method addresses these challenges using neural architecture search (discussed further in Section~\ref{sec:expnas}), achieving an average accuracy of 94.1\%, representing up to a 10.2\% improvement over existing approaches.

Table~\ref{tab:results} records the detection performances using the above-mentioned metrics. We compare three representative baselines (DRL, Ratafia, and Rwarmor) with our proposed method across various ransomware categories: Crypto, Locker, Scareware/FakeAV, and Hybrid. The final row shows the averages over all categories.

Overall, DRL reaches 78.4\% accuracy, Ratafia achieves 83.2\%, and Rwarmor falls behind at 76.1\%. Our proposed method significantly outperforms these approaches, with 91.9\% accuracy on average and improvements of 8–12\% across the different categories. Precision and recall follow the same trend: while DRL and Rwarmor often fall below 80\%, our method maintains values above 91\% in all categories, leading to the highest F1 scores. This indicates that our framework achieves both low false positives and low false negatives consistently.

The relative performance trends are consistent with the characteristics of each detection method. DRL is a static analysis-based method that relies solely on static features. Consequently, it struggles to differentiate between benign encryption programs and ransomware, causing the highest number of false alarms, resulting in low precision. Ratafia leverages dynamic monitoring of HPC events but lacks complementary ETB information, which results in higher FNRs when ransomware families exhibit subtle or obfuscated runtime behaviors, lowering its recall. Rwarmor also adopts dynamic analysis, but its feature recognition algorithm depends heavily on communication patterns, which became fragile in front of categories such as Scareware and Hybrid, where behaviors are less uniform. For instance, in Scareware/FakeAV, Rwarmor's accuracy drops to 70.8\%, demonstrating the difficulty of detecting ransomware that primarily relies on deceptive user-facing alerts rather than heavy encryption workloads. In contrast, our method integrates both HPC and ETB features and leverages NAS to optimize the detection model. This combination captures both coarse-grained runtime patterns and fine-grained instruction-level traces to ensure robust precision and recall. This improvement is especially significant when detecting Crypto and Scareware, where the proposed approach achieves over 93\% accuracy compared to 85.2\% and 72.1\% for DRL, respectively. These results confirm that the proposed method generalizes well across diverse ransomware behaviors and addresses the weaknesses of existing techniques.

%  DRL, as a static analysis framework, is prone to higher false alarms, leading to the largest FPR among all methods. Ratafia leverages dynamic monitoring of HPC events but lacks complementary ETB information, which results in higher FNRs when ransomware families exhibit subtle or obfuscated runtime behaviors. Rwarmor, which extends dynamic analysis but depends heavily on communication patterns, struggles in categories such as Scareware and Hybrid where behaviors are less uniform. For instance, in Scareware/FakeAV, its accuracy drops to 70.8\%, demonstrating the difficulty of detecting ransomware that primarily relies on deceptive user-facing alerts rather than heavy encryption workloads.

% Our proposed approach integrates both HPC and ETB features and employs neural architecture search to identify more effective model structures. This design enables the system to capture both coarse-grained execution patterns and fine-grained instruction traces, resulting in stronger discrimination between benign and malicious behaviors. The benefits are particularly visible in the Crypto and Scareware categories, where accuracy exceeds 93\% compared to 85.2\% and 72.1\% for DRL, respectively. Overall, the results confirm that the proposed method achieves robust detection across diverse ransomware types, addressing the shortcomings of purely static or dynamic techniques.

\subsection{Case study: Adversarial Robustness}
\label{sec:ablation_adv}

\begin{table*}[t]
\centering
\caption{Robustness to adversarial obfuscation, with improvements of the proposed method over baselines.}
\label{tab:adv_results}
\setlength{\tabcolsep}{5pt}
\footnotesize
\begin{tabular}{|c|cc|cc|cc|cc|ccc|}
\hline
\textbf{Category}
 & \multicolumn{2}{c|}{\textbf{DRL}}
 & \multicolumn{2}{c|}{\textbf{Ratafia}}
 & \multicolumn{2}{c|}{\textbf{Rwarmor}}
 & \multicolumn{2}{c|}{\textbf{Proposed}}
 & \multicolumn{3}{c|}{\textbf{Improvement (Proposed $-$ Baseline)}} \\
\cline{2-12}
 & Clean & Adv
 & Clean & Adv
 & Clean & Adv
 & Clean & Adv
 & vs DRL & vs Ratafia & vs Rwarmor \\
\hline
Crypto        & 85.2 & 65.3 & 88.9 & 74.2 & 81.3 & 70.5 & 93.8 & 89.6 & +24.3 & +15.4 & +19.1 \\
Locker        & 76.5 & 58.4 & 80.2 & 68.7 & 74.9 & 63.8 & 91.7 & 87.9 & +29.5 & +19.2 & +24.1 \\
Scareware/FAV & 72.1 & 52.0 & 79.4 & 66.5 & 70.8 & 59.7 & 89.9 & 85.7 & +33.7 & +19.2 & +26.0 \\
Hybrid/Misc   & 79.6 & 62.7 & 84.3 & 73.1 & 77.5 & 66.9 & 92.3 & 88.1 & +25.4 & +15.0 & +21.2 \\
\hline
\textbf{Average}
              & 78.4 & 59.6 & 83.2 & 70.6 & 76.1 & 65.2 & 91.9 & 87.8 & +28.2 & +17.2 & +22.6 \\
\hline
\end{tabular}
\end{table*}

Robustness is central to ransomware detection, as real attacks routinely use obfuscations to evade conventional signature-based detection. To evaluate the robustness of selected methods, we generate adversarial variants that perturb with the following five strategies:

\begin{itemize}
  \setlength{\itemsep}{2pt}
  \setlength{\parskip}{0pt}
  \setlength{\parsep}{0pt}
  \item \emph{NOP:} insert short runs of no-ops or benign ops to pad code without changing control flow.
  \item \emph{Opcode substitution:} swap instructions with semantics-equivalent forms to alter tokens and $n$-grams.
  \item \emph{Sleep:} inject brief sleeps between phases to create idle windows in the trace.
  \item \emph{API reordering:} permute independent API calls within a stage to disrupt sequence patterns.
  \item \emph{Packing:} wrap binaries with a simple packer for polymorphism.
\end{itemize}

Table~\ref{tab:adv_results} compares the clean and adversarial accuracy for all four methods. As we can see, DRL shows the steepest decline under obfuscation, falling from 78.4\% to 59.9\% on average and dropping below 60\% on Scareware, which is no better than a random guess. Rwarmor exhibits a similar pattern, with adversarial accuracy near 59\% on Scareware, while Ratafia holds up better overall but still slips to mid 60\% on Locker when sleeps and API reorderings disturb its HPC signatures. These outcomes align with their underlying properties: DRL focuses on static signatures which can be easily perturbed; Rwarmor relies on sequence regularities which are fragile against reordering; and Ratafia lacks ETB traces to anchor timing shifts. In contrast, our approach maintains 88.3\% on adversarial samples (-4.1\% from clean), with an average 22.6\% improvement over baselines, and achieved the largest gap on Scareware detection(34.4\%).

\begin{figure}[htbp]
\centering
\vspace{-0.15in}
\includegraphics[width = 0.45\textwidth]{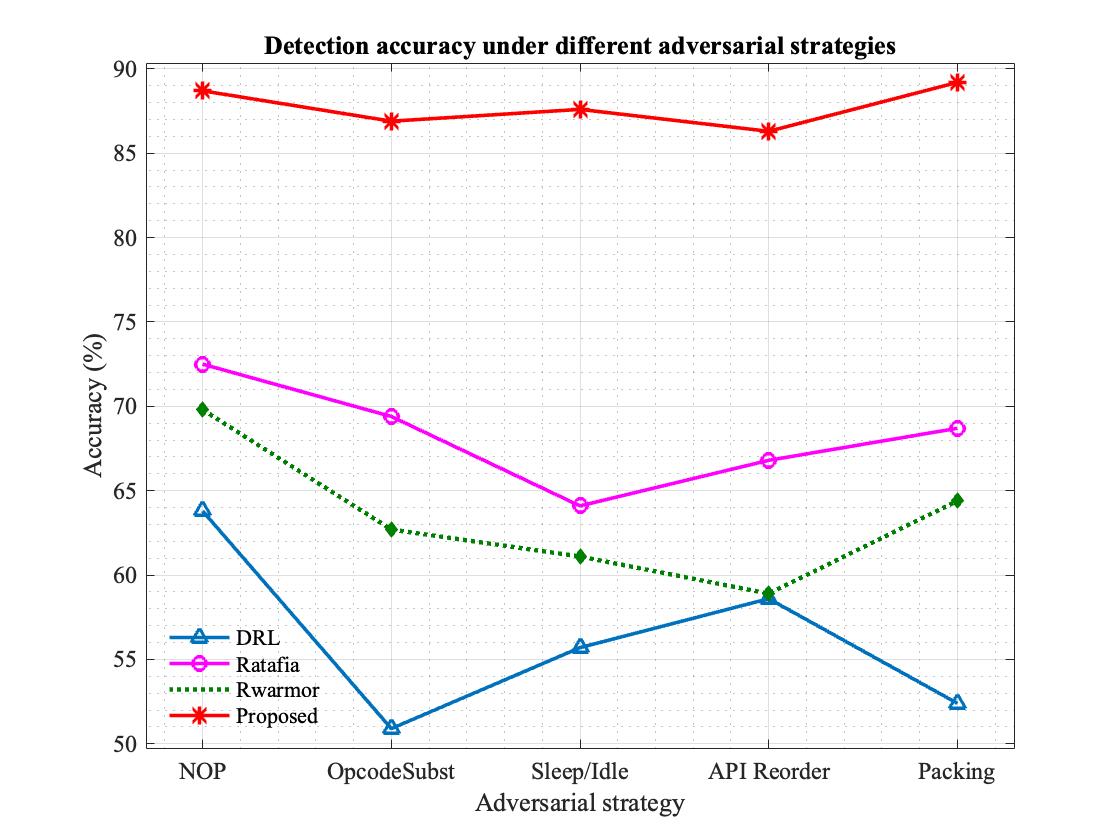}
\vspace{-0.1in}
\caption{The detection accuracy against adversarial strategies.}
\vspace{-0.2in}
\label{fig:adv_strategy}
\end{figure}

Figure~\ref{fig:adv_strategy} breaks down performance by adversarial strategies. Among the baselines, DRL is hit hardest by opcode substitution and packing, where tokens and structure-level signatures are muted; in both cases its accuracy falls below 50\%. Ratafia degrades most with sleep/idle frames insertion, as idle gaps blur the timing patterns in HPC traces, leading to accuracy down to mid–60\%. Rwarmor drops significantly under API reordering, where the control-flow features become messy and less informative, resulting in an accuracy around 59\%. In contrast, our approach remains steady across all five strategies, typically between 86\% and 90\%.
We attribute this advantage to two design choices. First, we adopt an adversarial training step for the static analysis, enhancing its robustness against such types of obfuscation during the training phase. Second, ETB-based tracing complements HPC-based tracing by providing fine-grained context. In this case, even obfuscation perturbs aggregate HPC statistics (e.g., idle windows or delayed activation), the underlying malicious activity with characteristic ETB patterns can still be captured in subsequent windows.

\subsection{Case study: Detection Latency}
Another critical performance metric is detection latency, which is the time elapsed from the initiation of the malicious behavior of ransomware to its detection. This latency is directly proportional to the number of files affected before discovery. Table~\ref{tab:latency} records the average detection latency across all tested methods. As indicated, our approach outperforms the others. Specifically, DRL exhibits the longest latency as it involves heavy computational overhead required by its reinforcement learning framework. Ratafia demonstrates slightly better efficiency, while our method achieves the lowest latency. This improvement is attributed to the sliding-window monitoring with hardware assistance.

\begin{figure}[htbp]
\centering
%\vspace{-0.2in}
\includegraphics[width = 0.32\textwidth]{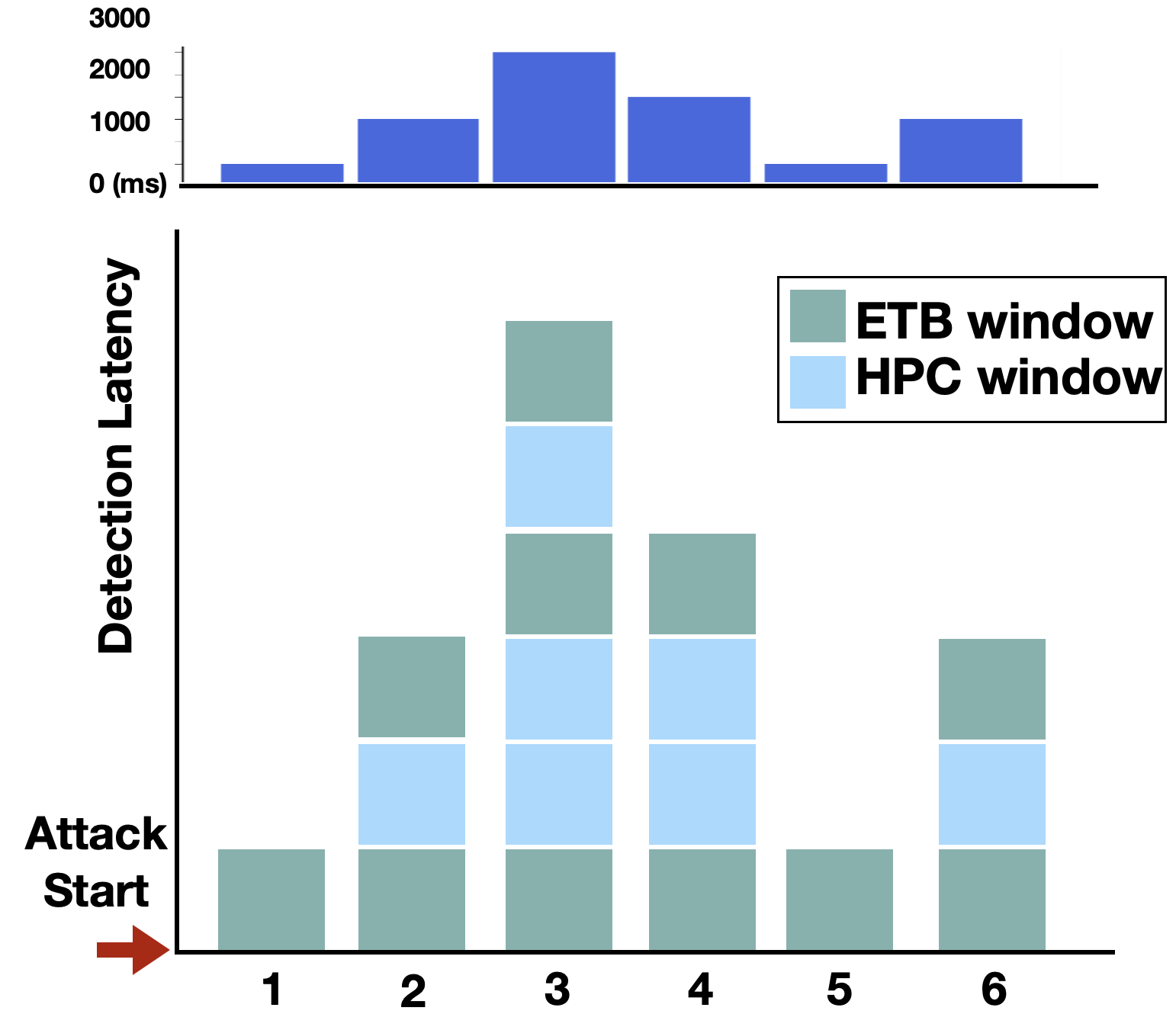}
\caption{Detection latency of six ransomware samples under sliding-window dynamic analysis with HPC and ETB.}
%\vspace{-0.2in}
\label{fig:latency}
\end{figure}

To illustrate, Figure~\ref{fig:latency} shows an in-depth inspection of six sample detection processes. In the lower diagram, the window-based monitoring progression is depicted, while the upper histogram tracks the total detection time for each sample. For example, the detector raises an alarm in the first window via initial ETB confirmation for sample 1 and 5. While for variants with delayed activation, such as sample 3, the system still identifies the threat within a small number of windows using the combined HPC–ETB pipeline as described in Section~\ref{sec:hardware}.

\begin{table}[htp]
\caption{The comparison of detection latency.}
\centering
\begin{tabular}{|c|clll|clll|clll|c|c|}
\hline
\textbf{Methods}   & \multicolumn{4}{c|}{\textbf{DRL}} & \multicolumn{4}{c|}{\textbf{Ratafia}} & \multicolumn{4}{c|}{\textbf{Rwarmor}} & \textbf{Proposed} & Improv \\ \hline
\textbf{Delay(ms)} & \multicolumn{4}{c|}{49984}        & \multicolumn{4}{c|}{13480}            & \multicolumn{4}{c|}{29670}            & 2784              & 4.8x   \\ \hline
\end{tabular}
\label{tab:latency}
\end{table}

\begin{figure}[t]
\centering
%\vspace{-0.2in}
\includegraphics[width=\linewidth]{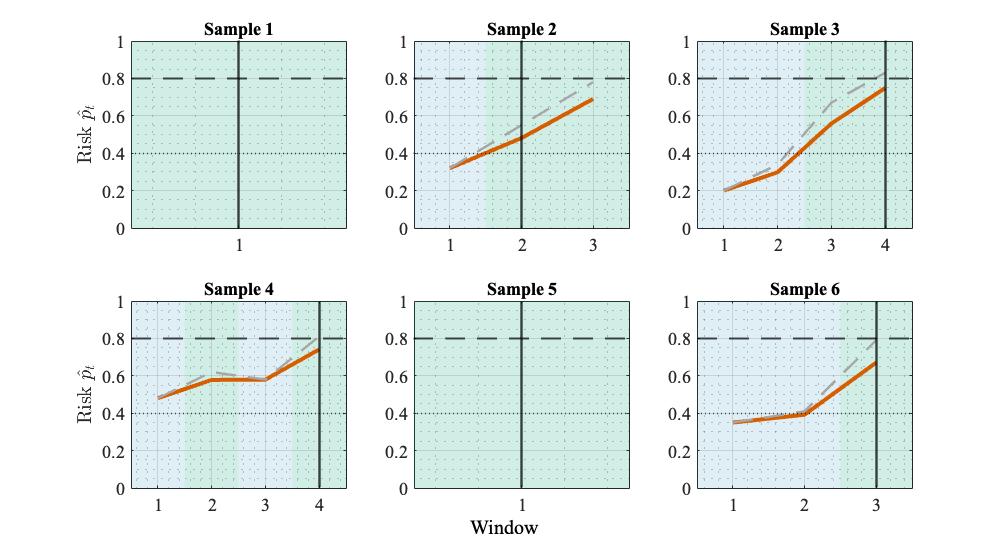}
\vspace{2pt}
{\footnotesize{(a) Smoothed risk $\hat{p}_t$ over 500\,ms windows for six samples.
Shading denotes monitoring mode (HPC vs.\ ETB); the vertical line marks the alarm window.}}

\medskip

\includegraphics[width=\linewidth]{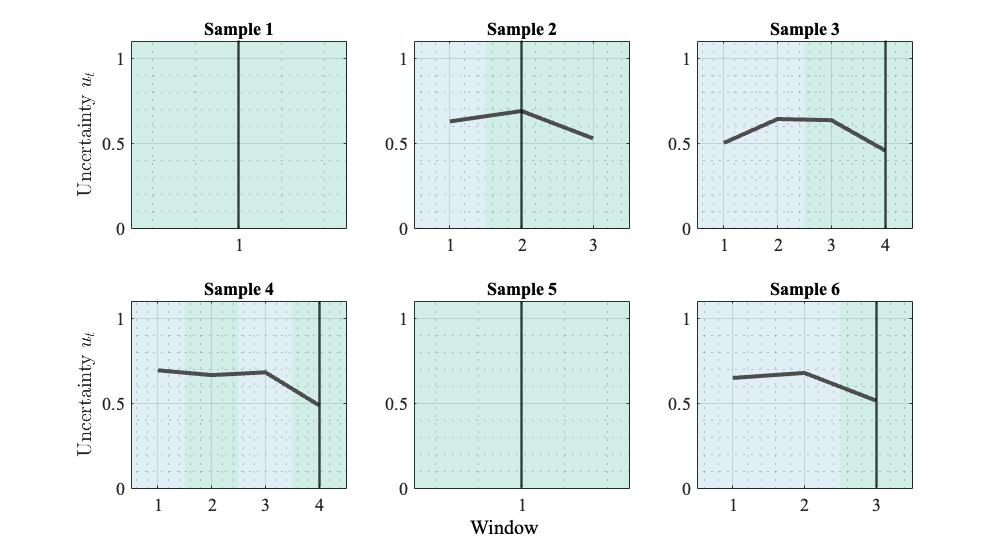}
\vspace{2pt}
{\footnotesize{(b) Predictive uncertainty $u_t$ for the same samples, aligned with (a).}}

\caption{Risk and uncertainty traces aligned with the six detection processes in Fig.~\ref{fig:latency}.}
\vspace{-0.1in}
\label{fig:risk_uncert}
\end{figure}

%\textbf{Risk and uncertainty over time.}
The reason behind these immediate responses is our risk-guided monitoring. Figure~\ref{fig:risk_uncert} plots the smoothed risk $\hat{p}_t$ and predictive uncertainty $u_t$ across the monitoring process for these samples. As shown, the risks for samples 1 and 5 exceed the threshold (set to 0.8 in our experiment) in the first window with low uncertainty, leading to immediate alarms. Sample 3 exhibits a delayed start: risk stays modest for two windows, then rises quickly, triggering a short ETB burst that exposes malicious activity; the alarm follows in the next window. Sample 4 shows a brief escalation and de-escalation: ETB is triggered once, but no malicious activity is confirmed, and the monitor returns to HPC without thrashing, thereby avoiding extra overhead. Overall, the proposed method delivers a fast and accurate response to ransomware activities.

% The reason behind such immediate responses to ransomware activities is our proposed risk-guided monitoring. Figures~\ref{fig:risk_uncert} plot the smoothed risk $\hat{p}_t$ and predictive uncertainty $u_t$ across the monitoring process for these samples. As we can see, the risks for samples 1 and 5 exceed the threshold (set to 0.8 in our experiment)in the first window with low uncertainty, leading to immediate alarms. Sample 3 exhibits a delayed start, where risk stays modest for two windows. But it rises quickly, causing one short ETB burst which exposes its activity, and the alarm follows in the next window. Sample 4 shows a brief escalation and de-escalation: ETB is triggered once, but no malicious activity is confirmed. The monitor returns to HPC without thrashing to avoid extra overhead. Overall, the proposed method presents an acute and accurate response to ransomware activities. 

% Two subfigures side-by-side without extra packages

\subsection{Case Study: Ablation Test of Dynamic Monitoring}
\begin{table*}[t]
\centering
\caption{Ablation of dynamic monitoring strategies: adversarial detection accuracy and detection latency (ms) by category.}
\label{tab:ablation_category}
\setlength{\tabcolsep}{5pt}
\footnotesize
\begin{tabular}{|l|c|c|c|c|c|c|c|c|}
\hline
\textbf{Category} &
\textbf{HPC-only Acc} &
\textbf{HPC-only Lat.} &
\textbf{ETB-only Acc} &
\textbf{ETB-only Lat.} &
\textbf{Adaptive Acc} &
\textbf{Adaptive Lat.} &
\textbf{Acc gain (pp)} &
\textbf{Latency gap} \\
\hline
Crypto
 & 74.6 & 5600
 & 90.8 & 2680
 & \textbf{89.1} & \textbf{3900}
 & 14.5 & 1220 \\

Locker
 & 65.8 & 6200
 & 88.9 & 2850
 & \textbf{87.4} & \textbf{4200}
 & 21.6 & 1350 \\

Scareware/FAV
 & 68.9 & 6400
 & 87.2 & 3000
 & \textbf{86.3} & \textbf{4500}
 & 17.4 & 1500 \\

Hybrid/Misc
 & 73.7 & 5800
 & 91.5 & 2600
 & \textbf{90.4} & \textbf{4000}
 & 16.7 & 1400 \\

\hline
\textbf{Average}
 & 70.8 & 6000
 & 89.6 & 2784
 & \textbf{88.3} & \textbf{4150}
 & \textbf{17.5} & \textbf{1367.5} \\
\hline
\end{tabular}
\end{table*}

We further conducted an ablation study to isolate the impact of different dynamic monitoring strategies, including an HPC-only variant, a PT-only variant, and the proposed adaptive monitoring policy that combines HPC with selective PT escalation. The results are summarized in Table~\ref{tab:ablation_category}.
Overall, the HPC-only variant exhibits the lowest detection accuracy, with an average drop of 17.5~percentage points compared to the adaptive policy, and incurs substantially higher detection latency. This behavior is particularly obvious for the \emph{Locker} and \emph{Scareware/FAV} categories, which are more staggered and delayed, causing the coarse-grained HPC-based detector to wait for additional window frames before reaching high confidence.
In contrast, the ETB-only variant achieves higher accuracy and the lowest detection latency by leveraging fine-grained execution traces, but at the cost of significantly increased runtime memory overhead due to continuous fine-grained tracing, making it expensive to maintain in practice. Instead, the proposed adaptive policy effectively balances this trade-off: it attains detection accuracy close to the PT-only configuration while maintaining latency comparable to, or lower than, the HPC-only baseline, which highlights the effectiveness of the risk-guided escalation and early-exit mechanism. As we can see, by escalating to ETB tracing only when needed, the adaptive framework reduces unnecessary cost for fine-grained monitoring while preserving fast detection in practice.

\subsection{Case study: Generalizability}
\label{sec:expnas}
\begin{figure}[htbp]
\centering
\vspace{-0.1in}
\includegraphics[width = 0.45\textwidth]{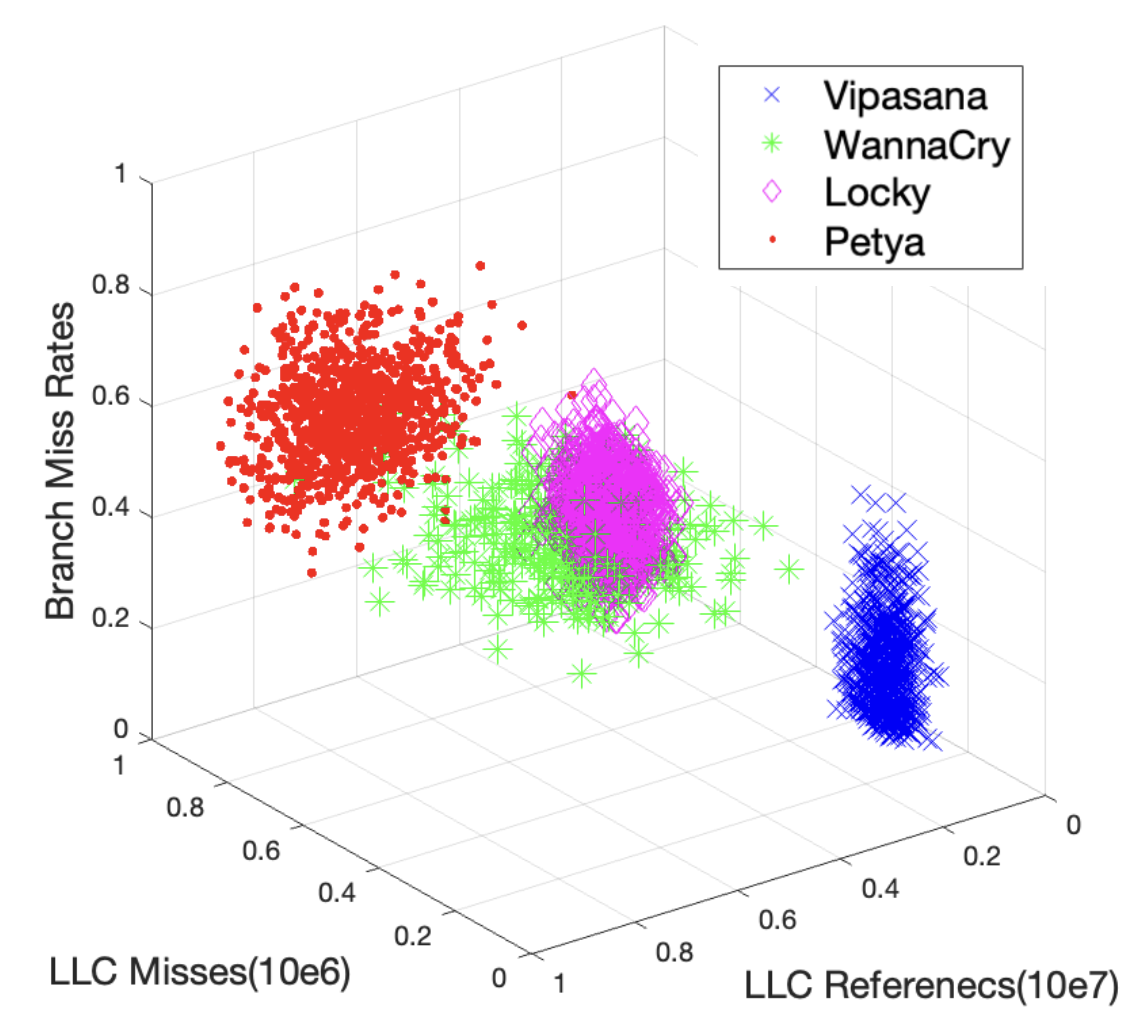}
%\vspace{-0.1in}
\caption{The scatter plot of selected HPC values for various benchmarks.}
%\vspace{-0.1in}
\label{fig:scatter}
\end{figure}

In this section, we take an in-depth inspection of how our proposed NAS-guided architecture addressed the generalizability problem in ransomware detection. As previously presented in Figure~\ref{fig:unstable} and Table~\ref{tab:results}, existing detection approaches yield unstable performance across different ransomware variants. To understand the reason, we plot HPC feature values of several selected ransomware samples in 3-D space as presented in Figure~\ref{fig:scatter}. As we can see from the scatter plot, a distinct observation is that different ransomware possess different feature clusters. In other words, there exist significant, distinct feature patterns for different ransomware variants. For example, as previously mentioned in Section~\ref{sec:results}, Vipasan possesses completely different patterns in the feature space, with low branch misses and low-level cache references due to the lack of communication with the control server. As a result, the designed ML model must possess flexibility so that it can capture various types of feature patterns. But it would be very hard to design such a machine learning model architecture manually, so that it could adapt to cover all variants.

\begin{figure}[htbp]
\centering
%\vspace{-0.1in}
\includegraphics[width = 0.5\textwidth]{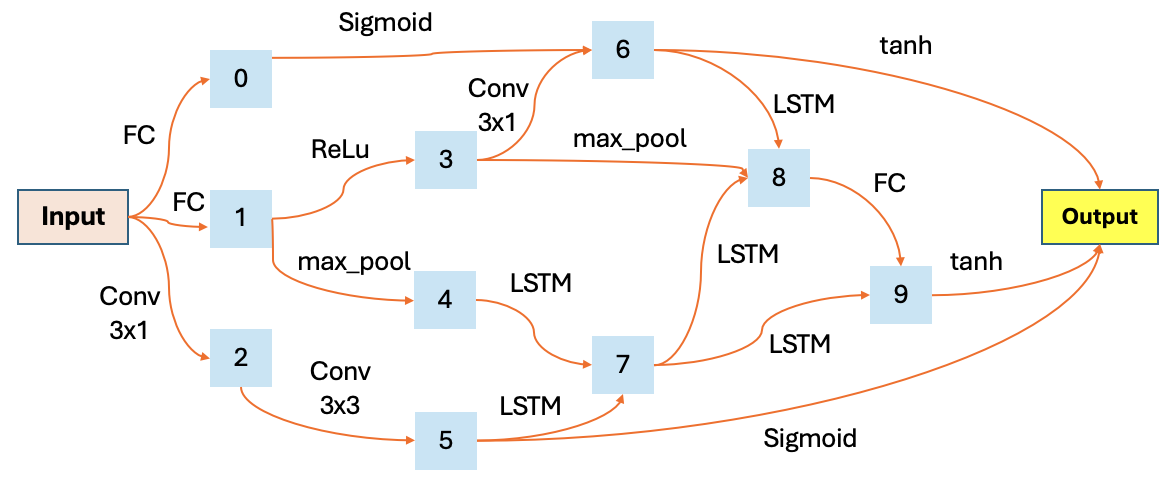}
%\vspace{-0.1in}
\caption{A simplified demonstration of our LSTM model generated from our neural architecture search (NAS) algorithm.}
%\vspace{-0.1in}
\label{fig:samplenas}
\end{figure}

Our proposed method applies neural architecture search (NAS) to address this challenge. Figure~\ref{fig:samplenas} depicts the simplified outline of our model generated from NAS. We interpret the workflow of this generated model in this way: after receiving the input values, blocks 0-2 are immediately utilized. These blocks serve as a ``classifier'' to firstly determine what type of variant it is, then different subsequent blocks and layers are applied after the ``classification'' step, so that different modules can be assigned to capture different types of feature patterns subsequently, therefore successfully perform detection across various ransomware variants. Though the rationale is straightforward and we could also manually design such an architecture, but it requires huge amount of tries and prior-knowledge. In our case, this architecture is generated automatically without costing any extra effort.

\subsection{Case Study: Explainable Machine Learning Analysis}
\label{sec:xaiexp}
In this section, we perform explainable machine learning analysis on the proposed method's outcome, to provide both feature-level and trace-level evidence for each model decision, increasing its transparency.

\begin{figure}[htp]
\centering
\includegraphics[width=\linewidth]{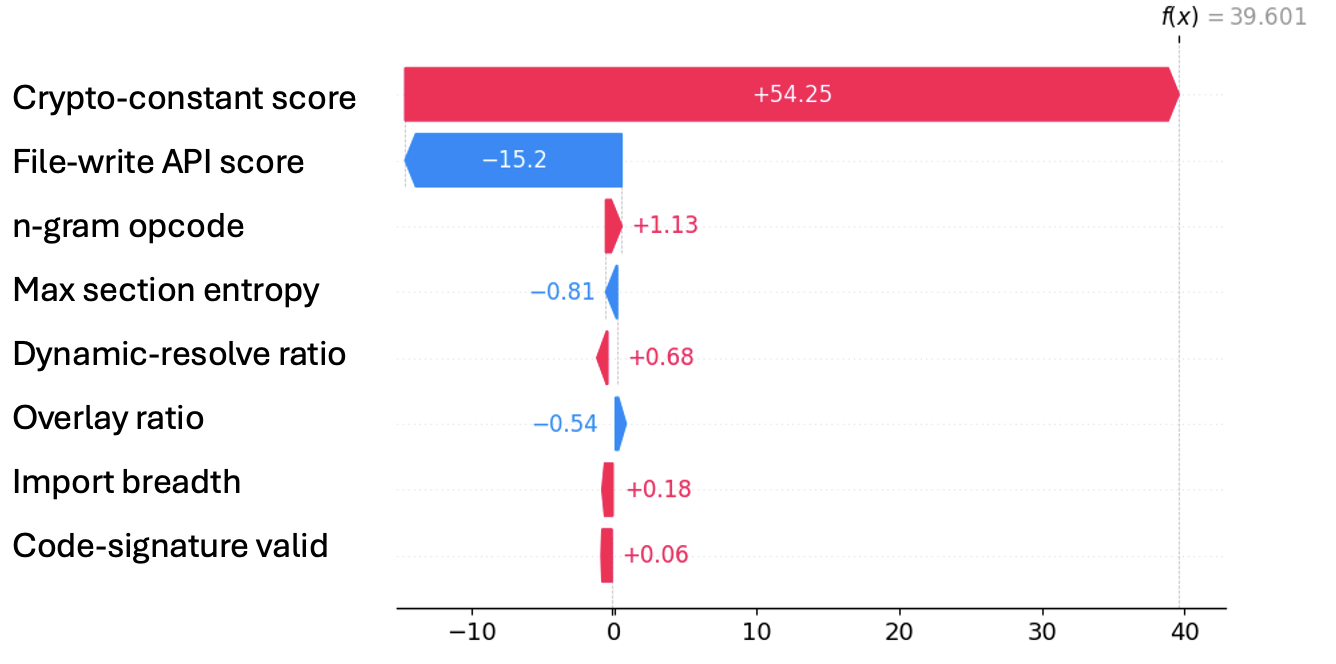}\\[4pt]
{\footnotesize{(a) True positive (Crypto).}}
\medskip
% (b) image, cropped a bit on the left
%\includegraphics[width=0.9\linewidth,trim=5pt 0pt 0pt 0pt,clip]{figures/SHAP2.png}
% (b) image, nudged left a touch
\vspace{0.1 in}
\hspace*{-10mm}\includegraphics[width=0.85\linewidth]{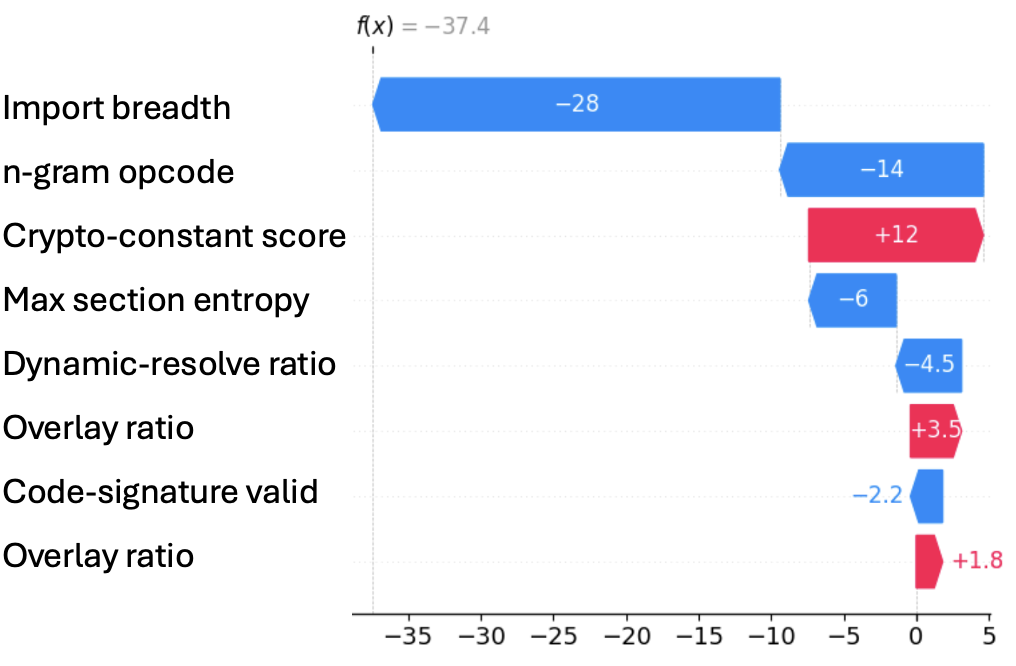}

{\footnotesize{(b) True negative (benign encryption). }}
\caption{SHAP waterfalls for two representative samples. Bars show per-feature contributions on the logit scale; the sum plus the base value gives $f(x)$, compared against the zero boundary.}
\label{fig:shap_pair}
\vspace{-0.15 in}
\end{figure}

First, as illustrated in Section~\ref{sec:xai}, we perform SHAP analysis on static analysis results. Figure~\ref{fig:shap_pair} shows waterfall plots of SHAP values for two representative samples: (a) a true positive {Crypto} case and (b) a true negative benign encryption program. We plot contributions on the model’s features, with the decision boundary at zero; red bars push the score toward ransomware, blue bars push it toward benign. Only the top eight features are shown for readability.
In Figure~\ref{fig:shap_pair} (a), the decision is dominated by \emph{Crypto-constant score}, while \emph{Import Breadth} and \emph{$n$-grams opcode} provide side support. A few features do push back, but their magnitudes are modest, so the model's final output ($f(x)$) lands well above the boundary, consistent with the detection outcome of this Crypto-type ransomware sample.

Instead, Figure~\ref{fig:shap_pair} (b) illustrates the opposite outcome for a benign program that legitimately uses cryptography. Although a crypto signal is present, the other stronger features dominate. 
A broad, regular \emph{import breadth} and a benign \emph{$n$-gram opcode} profile contribute the largest negative shifts. Low reliance on \emph{dynamic-resolve} and only moderate \emph{section entropy} provide additional evidence, and the \emph{valid code signature} reinforces the benign prior, leaving the final score well below the decision boundary.

% For example, a regular import profile with broad coverage, the absence (or very low level) of dynamic API resolution and packing artifacts, moderate section entropy, and a valid publisher signature. Collectively, these features pull $f(x)$ below zero, yielding a correct benign decision despite the presence of encryption.

Collectively, these two examples show that our proposed framework does more than return a label at the static stage. It also reports per-feature contributions, making the decision path explicit. Analysts can verify whether the dominant signals align with the sample’s nature (e.g., crypto intent for encryptors) for more accurate detection. Meanwhile, this breakdown can also be used to debug why an alarm was raised and to decide whether proceed to dynamic monitoring stage.

% These two examples highlight that the model does not key off a single cue; instead it weighs crypto intent alongside file-impact, loader, and layout signals. The waterfalls make that trade-off explicit and show why, in one case, the score crosses into the ransomware region while, in the other, benign priors prevail.

As for dynamic monitoring, we complement the explainability view with a local, LIME-based analysis over the runtime trace table. Specifically, we record the register/buffer values captured by HPC/ETB during a focused window, and fit a lightweight linear surrogate, where the resulting coefficients could be viewed as attributions.

\begin{figure}[htp]
\centering
% --- (a) Temporal view: columns as features ---
\includegraphics[width=0.7\linewidth]{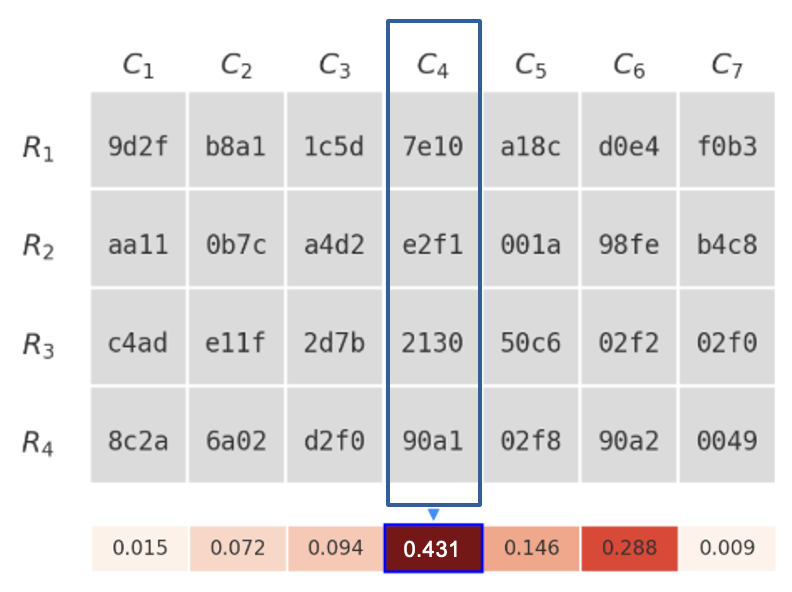}\\[4pt]
{\footnotesize {(a) Temporal view (columns as features). The largest coefficient at $C_4$ identifies the critical window driving the alert.}}

\medskip

% --- (b) Spatial view: rows as features ---
% Use trim/clip if you need to nudge alignment (left-crop a little).
\includegraphics[width=0.8\linewidth]{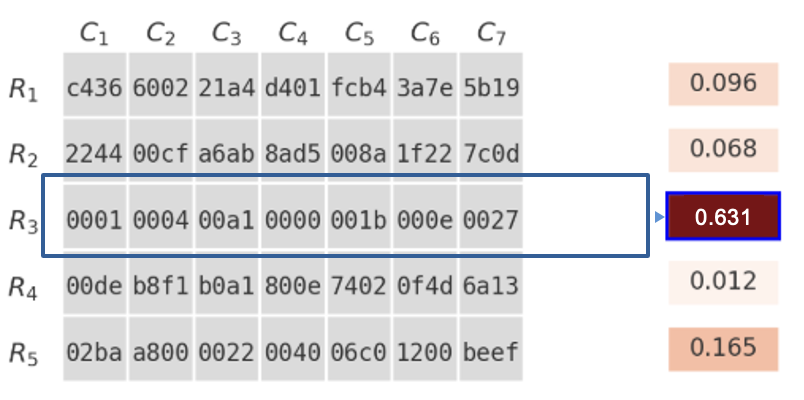}\\[4pt]
{\footnotesize {(b) Spatial view (rows as features). $R_3$ carries the dominant weight, localizing the anomalous register/buffer.}}

\caption{LIME-based local explanations on the runtime trace matrix. Panel (a) ranks time windows (when); panel (b) ranks channels (where). Coefficients are from a local linear surrogate fit to detector scores under small perturbations.}
\label{fig:lime_pair}
\end{figure}

Figure~\ref{fig:lime_pair} presents two analyst-facing views of a monitored ransomware sample. In Figure~\ref{fig:lime_pair} (a), we treat \emph{columns as features} and rank time slots. The largest weight lands on \(C_4\) (coefficient \(0.431\)), marking this clock cycle that drives the alert. This pinpoints \emph{when} the suspicious burst occurs and can be used to confirm the risk spike that triggered ETB escalation. In Figure~\ref{fig:lime_pair} (b), we flip the perspective and treat \emph{rows as features}; the surrogate highlights \(R_3\) (coefficient \(0.631\)) as the dominant contributor, indicating \emph{where} the malicious behavior happened, and which register/buffer was affected.

Together, these views provide human-understandable evidence to interpret the detection process. In this way, an analyst can verify the correctness of detection and trace back to the raw ETB/HPC signals for ransomware localization. The temporal–spatial attributions also help separate brief obfuscation from sustained malicious activity, reducing false positives and guiding when to terminate the target program. 
%The same evidence trail supports post-incident forensics and tuning of escalation thresholds, since the specific windows and channels that mattered are explicit.

% Together, these views provide human-understandable evidence to interpret the detection process. In this way, an analyst can verify the correctness of detection and can trace back to the raw ETB/HPC signals for ransomware localization.

\subsection{Case Study: Overhead Analysis}
\label{sec:overhead}

% We evaluate the compute and runtime costs of all four methods on the same host machine (i9@3.7\,GHz, RTX~3080, 64\,GB). Table~\ref{tab:overhead} summarizes training and inference overhead. The proposed NAS search is an \emph{offline} step (once per dataset/config) and adds \(\sim\)5.2\,h, after which training time is comparable to baselines. 

% As Table~\ref{tab:overhead} shows, the NAS search adds a one-time training cost. It remains well below the reinforcement-learning (DRL) budget, and while longer than Ratafia and Rwarmor, it is an offline step that runs once per configuration, so the impact is acceptable. At inference time, our model incurs no extra delay beyond the HPC/ETB pipeline. Additionally, all models fit comfortably under 250\,MB, making deployment practical on typical hosts.

To quantify the computational overhead of the proposed framework, we measure both offline and online costs under a controlled execution environment. Offline costs correspond to one-time expenses incurred during model preparation, including model training and NAS. Online overhead is measured during deployment and reported on a per-window basis, where each monitoring window spans 500\,ms of program execution. For online measurements, we instrument the runtime monitoring pipeline and record the wall-clock time consumed by each component. All runtime measurements were conducted on an otherwise idle system to minimize interference from background processes.
% HPC monitoring and decision aggregation are executed for every window, while PT tracing is triggered only when the adaptive policy escalates from coarse-grained monitoring to fine-grained tracing. 
% The reported average time reflects the mean cost across all windows, whereas the worst-case time corresponds to windows in which PT tracing is activated.

Table~\ref{tab:cost_breakdown} presents a component-level breakdown of these costs. 
As shown, though NAS is relatively time-consuming, it is an offline step that runs once per configuration, so the impact is acceptable. As for online components, HPC-based monitoring introduces a small and constant overhead, while ETB tracing incurs a higher per-window cost but remains bounded and infrequent. 
Consequently, the average end-to-end runtime overhead remains low, and the worst-case overhead is limited to the cost of a single ETB tracing window.

% \vspace{-0.1 in}
% \begin{table}[htp]
% \centering
% \caption{Compute/overhead comparison.}
% \label{tab:overhead}
% \setlength{\tabcolsep}{3.5pt}
% \footnotesize
% \begin{tabular}{|l|c|c|c|c|}
% \hline
% \textbf{Method} & \textbf{Train (h)} & \textbf{NAS (h)} & \textbf{Inference (ms)} & \textbf{Model (MB)} \\ \hline
% DRL              & 43.5  & -- & 5.2           & 226 \\ \hline
% Ratafia          & 13.1  & -- & 3.3           & 135  \\ \hline
% Rwarmor          & 16.5  & -- & 3.4           & 173  \\ \hline
% Proposed & 14.7  & 5.2  & 0.6 / 3.5    & 168  \\ \hline
% \end{tabular}
% \end{table}

\begin{table}[htp]
\centering
\caption{Cost breakdown of the proposed framework. Offline costs are one-time expenses for training and NAS, while online overheads are measured per monitoring window.}
%\vspace{-0.1 in}
\label{tab:cost_breakdown}
\setlength{\tabcolsep}{4pt}
\footnotesize
\begin{tabular}{|l|l|c|c|}
\hline
\textbf{Phase} &
\textbf{Component} &
\textbf{Avg. time} &
\textbf{Worst-case} \\
\hline
\multirow{3}{*}{Offline}
 & Static model training (DT)        & 2.1~h  & --  \\
 & Dynamic model training (LSTM)     & 12.6~h & --  \\
 & Neural architecture search (NAS)  & 5.2~h  & --  \\
\hline
\multirow{3}{*}{Online}
 & HPC monitoring                    & 0.6~ms & 0.8~ms \\
 & ETB tracing                        & 2.9~ms & 3.5~ms \\
 & Decision \& aggregation           & 0.2~ms & 0.3~ms \\
\hline
\multirow{2}{*}{Summary}
 & End-to-end (avg.)                 & \textbf{1.1~ms} & -- \\
 & End-to-end (worst)                & -- & \textbf{4.6~ms} \\
\hline
\end{tabular}
\end{table}

% \begin{table}[htp]
% \centering
% \caption{Cost breakdown of the proposed framework. Offline costs are one-time expenses for training and NAS; online overheads are measured per monitoring window during deployment. PT tracing is activated conditionally by the adaptive policy.}
% \label{tab:cost_breakdown}
% \setlength{\tabcolsep}{4pt}
% \footnotesize
% \begin{tabular}{|l|c|c|}
% \hline
% \textbf{Component} &
% \textbf{Avg. time} &
% \textbf{Worst-case} \\
% \hline
% \multicolumn{3}{|c|}{\textbf{Offline cost (one-time)}} \\
% \hline
% Static model training (DT)   & 2.1~h  & --    \\
% Dynamic model training (LSTM) & 12.6~h & --   \\
% Neural architecture search (NAS) & 5.2~h & --  \\
% \hline
% \multicolumn{3}{|c|}{\textbf{Online overhead (per 500\,ms window)}} \\
% \hline
% HPC monitoring               & 0.6~ms & 0.8~ms  \\
% PT tracing                   & 2.9~ms & 3.5~ms  \\
% Decision \& aggregation      & 0.2~ms & 0.3~ms  \\
% \hline
% \textbf{End-to-end (avg.)}   & \textbf{1.1~ms} & -- \\
% \textbf{End-to-end (worst)}  & -- & \textbf{4.6~ms}  \\
% \hline
% \end{tabular}
% \end{table}

\section{Conclusion}
\label{sec:con}

 Ransomware has become a significant threat to cybersecurity in recent years. Existing ransomware detection methods suffer from four major limitations: vulnerability to obfuscation, detection latency, and low generalizability. To address these challenges, this paper presents a dynamic hardware-assisted ransomware detection framework that leverages adaptive runtime monitoring to balance detection accuracy and efficiency. Experimental results demonstrate that the proposed approach achieves consistent improvements in detection accuracy and reduced detection latency compared to prior works. Additionally, by incorporating adversarial training, we strengthen static analysis against obfuscation techniques, and the utilization of one-shot Neural Architecture Search (NAS) automates the optimization of model architectures, enabling the framework's flexibility across ransomware variants.

\bibliographystyle{IEEEtran}
\bibliography{bibliography}


% \end{thebibliography}

\section*{Authors' Biographies}

\noindent\textbf{Salma Aboelmagd} (Graduate Student Member, IEEE) received the B.Sc. in Electrical and Computer Engineering from Texas A\&M University in 2020 and the M.Sc. in Data Science and Engineering from Hamad Bin Khalifa University in 2022. She is currently pursuing the Ph.D. degree in Electrical Engineering at Florida State University. 
%Her research focuses on data-driven cyber-attack detection in cyber-physical systems (with applications to UAVs and power systems), machine learning, and anomaly detection.

\noindent\textbf{Md. Rakibul Ahasan} (Graduate Student Member, IEEE) received the B.S. degree from Rajshahi University of Engineering and Technology, Bangladesh, in 2011, the M.S. degree from BRAC University, Bangladesh, in 2022, and another M.S. degree from Miami University, Ohio, in 2024. He is currently pursuing the Ph.D. degree in Electrical Engineering at Florida State University under Dr.~Abdulrahman Takiddin.
%His research interests include data-driven cyber-attack detection in cyber-physical systems, routing optimization, anomaly detection, and machine learning.

\noindent\textbf{Zhixin Pan} (Member, IEEE) received the Ph.D. degree in Electrical and Computer Engineering from the University of Florida in 2022. He is currently an Assistant Professor in the Department of Electrical and Computer Engineering at Florida State University. His research interests include hardware security, explainable artificial intelligence, and efficient machine learning systems.

\noindent\textbf{Abdulrahman Takiddin} (Member, IEEE) received the Ph.D. degree from Texas A\&M University in 2023. He is an Assistant Professor in the Department of Electrical and Computer Engineering at the FAMU–FSU College of Engineering, Florida State University. His research interests include machine learning, cyber-physical systems, smart grids, smart transportation, and security.

\noindent\textbf{Ziyu Shu} (Member, IEEE) is a Postdoctoral Researcher in the Department of Radiation Oncology at Washington University in St.~Louis. He received the Ph.D. degree in Computer Science from the University of Florida in 2023. His research interests include machine learning, medical imaging, and computer vision.

\end{document}